\documentclass[aps,twocolumn,superscriptaddress,footinbib,nobalancelastpage,prx]{revtex4-2}
\usepackage{amsmath,amssymb}
\usepackage[colorlinks=true,citecolor=blue]{hyperref}
\usepackage{graphicx}
\usepackage{amsmath}
\usepackage{slashed}

\usepackage[table,xcdraw,dvipsnames]{xcolor}

\newcommand{\be}{\begin{equation}}
\newcommand{\ee}{\end{equation}}

\newcommand{\ads}{{\rm AdS}}
\newcommand{\cft}{{\rm CFT}}
\newcommand{\Geff}{G^{\rm eff}}
\newcommand{\sbh}{S^{\ads_3 {\rm BH,simulator}}}

\begin{document}

\title{AdS/CFT Correspondence with a 3D Black Hole Simulator}

\author{Aydin Deger}
\affiliation{Department of Physics and Astronomy, University College London, London WC1E 6BT, United Kingdom}
\affiliation{School of Physics and Astronomy, University of Leeds, Leeds LS2 9JT, United Kingdom}

\author{Matthew D. Horner}
\affiliation{School of Physics and Astronomy, University of Leeds, Leeds LS2 9JT, United Kingdom}
\affiliation{Aegiq Ltd., Cooper Buildings, Arundel Street, Sheffield, S1 2NS, UK}
\author{Jiannis K. Pachos}
\affiliation{School of Physics and Astronomy, University of Leeds, Leeds LS2 9JT, United Kingdom}

\begin{abstract}
One of the key applications of AdS/CFT correspondence is the duality it dictates between the entanglement entropy of Anti-de Sitter (AdS) black holes and lower-dimensional conformal field theories (CFTs). Here we employ a square lattice of fermions with inhomogeneous tunneling couplings that simulate the effect rotationally symmetric 3D black holes have on Dirac fields. When applied to 3D BTZ black holes we identify the parametric regime where the theoretically predicted 2D CFT faithfully describes the black hole entanglement entropy. With the help of the universal simulator we further demonstrate that a large family of 3D black holes exhibit the same ground state entanglement entropy behavior as the BTZ black hole. The simplicity of our simulator enables direct numerical investigation of a wide variety of 3D black holes and the possibility to experimentally realize it with optical lattice technology.
\end{abstract}

\maketitle

\section{Introduction}

Spacetime geometry changes dramatically across the horizon of a black hole. Classical particles or even light that fall across the horizon can never escape, purely due to the structure of spacetime. Surprisingly, quantum correlations can be built across the black hole horizon, a phenomenon that leads to Hawking radiation \cite{Bekenstein1973,hawking_black_1974}. Conceptually, this mechanism is equivalent to quantum tunneling across a potential barrier \cite{Hawking:1975aa,Susskind:2006aa}. This phenomenon is not only confined to astronomical objects but can also be met in condensed matter or synthetic quantum systems. Recently, signatures of Hawking radiation have been identified in diverse systems, such as Bose-Einstein condensates \cite{Garay2000}, quantum Hall effect \cite{stone_analogue_2013}, Weyl fermions \cite{volovik_black_2016}, critical Floquet systems \cite{lapierre_emergent_2020}, magnons \cite{Magnonic2017} or chiral interfaces~\cite{Horner2023}.

It has been long hypothesized that the entanglement entropy of quantum fields in black hole geometry contributes to the Bekenstein-Hawking entropy, also known as the area law \cite{Bombelli1986,Srednicki1993,Cardy1986,kabat:1994aa,Strominger1996,Strominger1998,Aharony:2000aa,Emparan:2006aa,MAIELLA:2007aa,solodukhin:2011aa}. The area-law behavior also appears in Ryu--Takayanagi formula that enables holographic entanglement entropy calculations using AdS/CFT correspondence \cite{ryu_holographic_2006,ryu_aspects_2006,Rangamani_2017,Almheiri2021}.
This holographic principle provides a bridge between the theories of gravity in $D+1$ dimension with quantum field theories in $D$ dimension \cite{Maldacena1999,hooft:2009aa}. In particular, AdS/CFT emerges as a powerful tool for probing certain strongly coupled CFTs. For example, this correspondence has been used to discover new strongly coupled phenomena, non-Fermi liquids \cite{cubrovic:2009aa,liu:2011aa}. While the entanglement entropy of 2D systems is fairly well understood~\cite{calabrese_entanglement_2004,eisert_colloquium_2010}, the generalization to higher dimensions entails many subtleties~\cite{Casini_2005,eisert_colloquium_2010,nishioka_holographic_2009,casini_towards_2011,zhang_entanglement_2011}.

Here we present a quantum simulator of massless Dirac fermions in the gravitational background of black hole horizons. Our simulator is in three spacetime dimensions, though our approach can be effortlessly extended to higher dimensions. It has been shown that the radiation of black holes due to fluctuating gravity in the semiclassical limit is equivalent to the radiation of scalar or fermionic particles in the black hole background \cite{morsink_black_1991,Mann1991}. Hence, our black hole simulator can numerically and analytically probe static and dynamic properties of semi-classical quantum gravity that might be otherwise inaccessible. 

The simulator consists of a two-dimensional square lattice of fermions. By choosing the tunneling couplings of the lattice appropriately the system can be effectively described by Dirac fermions embedded in any black hole geometry~\cite{shi_chip_2021}. To test the validity of the simulator we employ the equivalence to the Unruh effect \cite{unruh1976} and show that the temperature of the black hole radiation is accurately described by Hawking temperature for a wide range of black hole profiles. Subsequently, we investigate the entanglement entropy of 3D black holes. We identify the parametric regime where the BTZ entanglement entropy numerically obtained from our 3D black hole simulator is in agreement with the theoretically predicted value of the corresponding 2D CFT that lives on the boundary of the AdS spacetime \cite{Brown1986,Aharony:2000aa,Emparan:2006aa,MAIELLA:2007aa,Cadoni2010,Khveshchenko_2013,Palumbo:2016aa}. 

Our work holds significance at both fundamental and practical levels. From a fundamental perspective, the proposed simulator offers a valuable tool to investigate quantum correlations of black holes, establishing a ``black hole laboratory" for exploring unresolved questions in gauge/gravity dualities. Our work also provides further supporting evidence for the conjecture that $\cft_2$ also describes various non-BTZ black hole profiles near the horizon, addressing the open problem of universality \cite{Carlip2002,Carlip:2007aa,Strominger:2009aa}.

In practical terms, our proposed simulator stands out as both simple and powerful, and it offers itself various generalizations. Additionally, it is based on a free theory, in contrast to corresponding conformal field theories that often involve interactions and thermal effects. This distinction introduces complexities when calculating entanglement entropy in higher dimensions. Consequently, our approach lays the foundation for investigating spatial correlations in interacting theories using a free theory in higher dimensions. Furthermore, our quantum simulator comprises a free fermion lattice that can be realized with many quantum technologies, such as cold atoms or Josephson junctions. This presents an exciting opportunity to simulate black hole physics in a laboratory setting \cite{Munoz-de-Nova:2019aa,Yanay:2020aa}.

\section{The model}

We now construct a universal simulator of 3D Dirac fermion in arbitrary black hole geometry. This simulator consists of a square lattice of fermions with position-dependent tunneling couplings. For simplicity, we employ a rotationally symmetric gravitational field with line element
\begin{equation}
ds^2 = F(r) d\tau^2 - F(r)^{-1} dr^2 - r^2d\theta^2,
\label{eq:metric}
\end{equation}
where $F(r)$ is a function only of the radial coordinate $r$. Dirac fermions with mass $m$ in the geometric background \eqref{eq:metric} satisfy
\be
i\slashed{\nabla} \Psi(\tau,r,\theta) = m\Psi(\tau,r,\theta),
\label{eq:Dirac1}
\ee
where $\slashed{\nabla} = e^\mu_a \gamma^a \partial_\mu + {1 \over 2|g|^{1/2}} \gamma^a \partial_\mu(|g|^{1/2} e^\mu_a) $. The dreibeins $e^\mu_a$ are defined by $g^{\mu \nu } = e^\mu_a e^\nu_b \eta^{ab}$, with $\eta^{ab}=\rm{diag}(1,-1,-1)$. The gamma matrices $\gamma^a$ satisfy the Clifford algebra $\{\gamma^a,\gamma^b\} = \eta^{ab} $. Due to the rotational symmetry of space \eqref{eq:metric} the spinor $\Psi$ can be written as $\Psi(\tau,r, \theta) = \psi(\tau,r)\chi(\theta)$, where $\slashed{\nabla}_\theta \chi(\theta) = \kappa \chi(\theta)$ and $\slashed{\nabla}_{\tau,r} \psi(\tau,r) = ({\kappa \over r} i\sigma^z -i m \mathbb{I}_2)\psi(\tau,r)$~\cite{lopez-ortega_dirac_2009}. The parameter $\kappa$ is a positive (non-zero) integer, corresponding to angular momentum eigenvalues \cite{camporesi_eigenfunctions_1996}. In the massless limit ($m \to 0$) and in the low energy regime ($\kappa$ small) the region with large $r$ is described by
\be
\slashed{\nabla}_{\tau,r} \psi (\tau,r) \approx 0.
\label{eq:2Dirac0}
\ee
This derivation can be generalized to higher dimensions. 

We now encode the 3D Dirac equation \eqref{eq:Dirac1} with black hole background in a simulator consisting of a square lattice of fermions. We employ a generalization of the procedure employed in~\cite{morsink_black_1991, yang_simulating_2020, pedernales_dirac_2018} for 2D black holes to the case of radially symmetric 3D black holes. To avoid coordinate singularity at the black hole horizon, we perform a change of variable $dt=d\tau+F(r)^{-1}dr$ and work in the ingoing Eddington-Finkelstein coordinates, $ds^2 = F(r) dt^2 - 2 dt dr$. We consider now the Dirac equation written in these coordinates. As the Dirac spinor in \eqref{eq:2Dirac0} is massless it can be written as $\psi(t,r) =\big[\phi(t,r),-\phi(t,r)\big]^T/\sqrt{2}$, i.e. the two components depend on each other so they do not need to be encoded independently in our lattice.
As a result \eqref{eq:2Dirac0} simplifies to
\begin{equation}
\partial_t \phi(t,r) = -\big[\partial_r (F(r) \phi(t,r)) + F(r) \partial_r \phi(t,r)\big]/4.
\label{eq:2Dirac1}
\end{equation}
The representation of \eqref{eq:2Dirac1} on a square lattice is obtained by discretizing the spatial position with a lattice constant $a$ (we fix $a=1$) and approximating the spatial derivatives with central differences. This is followed by substituting $\phi=\hat{f}$ where $\{\hat{f}_i,\hat{f}_j^\dagger\}=\delta_{ij}$ and using Heisenberg equation of motion $i\partial_t \hat{f}_j =[\hat{f}_j,\mathcal{H}]$, see Appendix \ref{app:dirac}. In the low energy limit where $\phi$ is smooth and for slowly varying functions $F(r)$, the resulting lattice system can be described by free fermions on a two-dimensional square lattice with nearest neighboring hopping
\be
\mathcal{H}=-\frac{1}{4}\sum_{\langle i,j \rangle}F_i \left(\hat{f}_{i}^\dagger \hat{f}_{j} + {\rm h.c}\right),
\label{eq:Ham}
\ee
where $F_i$ is the value of $F(r)$ with $r$ the polar distance of the vertex $i$ of the square lattice. The black hole geometry dictates that $F(r)$ in \eqref{eq:metric} turns from positive to negative as $r$ moves from outside to inside the black hole. The horizon is positioned at $r_h$ where $F(r_h)=0$. Due to the lattice nature of the simulator \eqref{eq:Ham} we can choose the couplings $F_i$ to never become zero everywhere around the circle with radius $r_h$. Nevertheless, the transition from positive to negative values of $F_i$ faithfully encodes the black hole spacetime geometry. As we demonstrate in the following, this simulator can faithfully describe the properties of Dirac fermions near a black hole horizon, taken to be at a large radius, where \eqref{eq:2Dirac0} is valid. 

Here we will begin with the BTZ black hole profile.
In the presence of negative cosmological constant $\Lambda = -1/l^2$, the most prominent solution to Einstein's equations is the three-dimensional locally $\ads_3$ BTZ black hole \cite{BTZ1992,Mann1997}. The metric of the BTZ black hole with mass $M$ is given by Eq.~\eqref{eq:metric} with $F^\text{BTZ}=(r^2-r_h^2)/l^2$. The horizon of the BTZ black hole is at position $r_h=2l\sqrt{2GM}$ and its Hawking temperature is given by $T_H=\sqrt{2 G M}/(l\pi)$. Next we will illustrate the numerical determination of the Hawking temperature for the BTZ black hole. Various other profiles will be considered in the last section.

\begin{figure}
  \includegraphics[width=\linewidth]{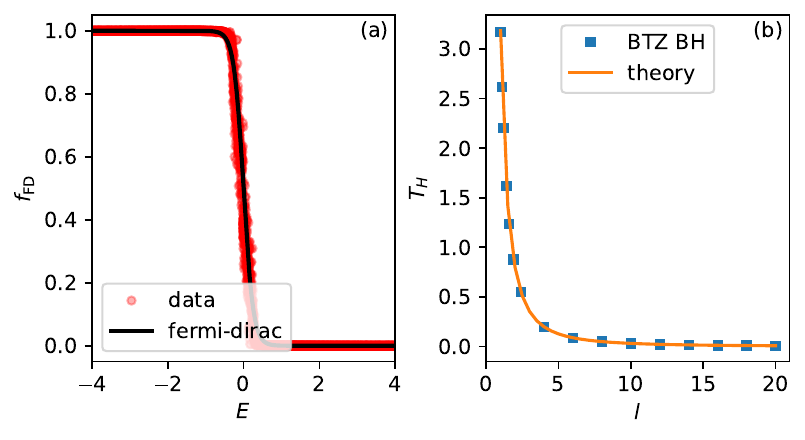}
\caption{Hawking temperature, $T_H$, determined from 3D black hole simulator for BTZ black hole with profile $F^{\rm BTZ}(r)=(r^2-r_h^2)/l^2$ (orange squares). (a) Data points are numerically measured for $l=5$ in a BTZ black hole profile. The solid line indicates the corresponding fit to a Fermi-Dirac distribution, $f_\text{FD}(E, T_H)$, where $T_H$ is extracted. (b) The measured temperature $T_H$ for a large range of parameter $l$ is in agreement with the theoretical Hawking temperature within 0.46\% error. Here we used system size $L=81$ and horizon radius $r_h=20$.}
\label{fig:prob}
\end{figure}  

\section{Hawking temperature}

We demonstrate now that our simulator can faithfully reproduce the theoretically predicted Hawking temperature of black holes, see Appendix \ref{app:hawking}. To determine the Hawking temperature from our black hole simulator we use the equivalence to the Unruh effect \cite{Mertens2022}. Black hole metrics are approximately equal to the Rindler metric close to the horizon that has a linear profile $F^{R}(r)= \eta(r-r_h)$. A stationary observer close to the black hole horizon can be equivalently described by a locally accelerating frame of reference moving through a flat Minkowski spacetime. Therefore, they will experience the Unruh effect with a temperature given by the Hawking temperature, $T_\mathrm{H}$. To simulate this effect we first encode Hamiltonian $\mathcal{H}_\mathrm{M}$ that describes Dirac fermions in local Minkowski frame, with many-body ground state $|0_\mathrm{M}\rangle$. We achieve that by taking a flat profile $F_i=F$ in our simulator of Eq.~\eqref{eq:Ham}. Then we simulate the local Rindler Hamiltonian which after diagonalization is given by $\mathcal{H}_\mathrm{R} = \sum_p E_p c^\dagger_p c_p$, with eigenmodes $\{ c_p \}$. Finally, the Rindler observer measures the mode occupation $\langle 0_\mathrm{M}|c^\dagger_p c_q |0_\mathrm{M}\rangle = f_\mathrm{FD}(E_p,T_\mathrm{H}) \delta(p - q)$, where $f_\mathrm{FD}(E_p,T_\mathrm{H})= (e^{E_p/T_H} +1)^{-1}$ is the Fermi-Dirac distribution at the Hawking temperature $T_\mathrm{H}$ and $E_p$'s are the single-particle energies of the Rindler Hamiltonian $\mathcal{H}_\mathrm{R}$, see Appendix \ref{app:unruh}. In the simulation, we take modes $p$ close to the ground state, where the continuum limit holds and determine the Fermi-Dirac distribution, as shown in Fig.~\ref{fig:prob}(a), from which we extract $T_H$. Repeating this process for various BTZ profiles we find that the simulator reproduces the theoretical predicted Hawking temperatures with remarkable accuracy with an error of $0.46\%$, as shown in Fig.~\ref{fig:prob}(b). We find that accuracy increases with lattice size.

The source of the resulting thermality is due to the fact that the Minkowski ground state exists on both sides of the horizon, whilst the local Rindler modes only have support outside of the horizon. Hence, projecting $|0_\mathrm{M}\rangle$ onto $\{ c_p \}$ effectively traces out the region inside the black hole, resulting in a thermal state. In the following, we will employ this 3D black hole simulator to investigate the entanglement entropy across the event horizon and compare it to the Bekenstein-Hawking entropy predicted. Through the AdS/CFT correspondence the Bekenstein-Hawking entropy can also be understood as the thermal entropy of the boundary CFT.

\begin{figure}
\includegraphics[width=\linewidth]{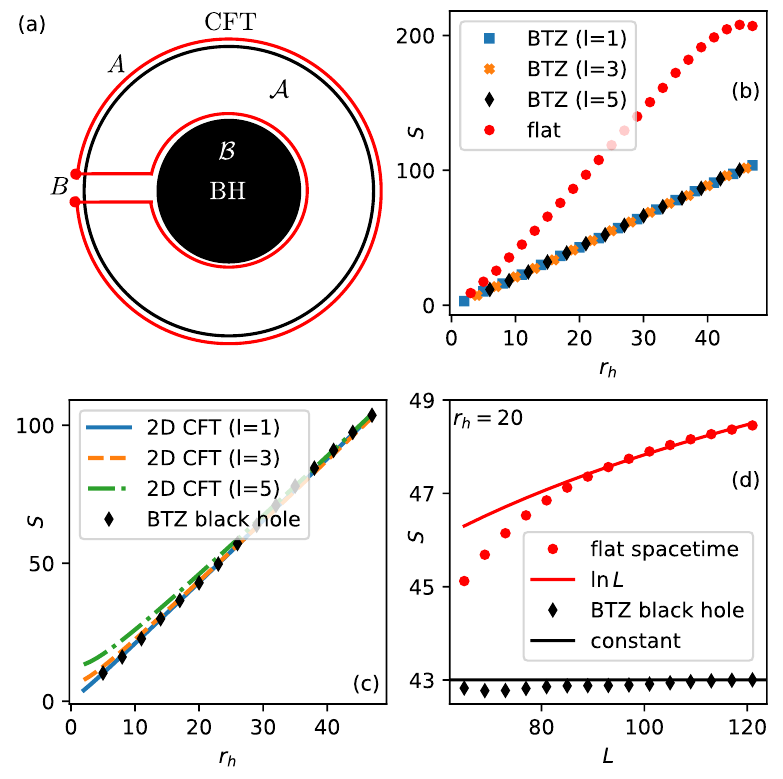}
\caption{(a) The holographic $\ads_3/\cft_2$ duality is illustrated for a black hole with path bipartitions the boundary in $A$ that covers the whole CFT and its trivial complement $B$ and wraps around the black hole horizon separating it into regions $\mathcal{A}$ and $\mathcal{B}$. (b) Red circles indicate the entanglement entropy of a flat spacetime whereas the rest of the colours correspond to BTZ black holes with different curvatures $l$. While $T_H$ changes as $l$ varies all of them have the same entanglement entropy across the horizon. The slope gives an effective Newton constant for BTZ black hole as $\Geff=0.7$.
(c) The CFT entropy for various curvatures $l$ (corresponding to different central charges) and the BTZ entanglement entropy. The holographic correspondence holds accurately in the large temperature limit $r_h\gg l$. 
(d) The entanglement entropy of flat spacetime (red, points) and the BTZ black hole (black, diamond) as a function of system size, $L$, with fixed radius $r_h=20$. The flat spacetime entanglement (value shifted by -44) scales with $\ln L$ (red, solid line) which indicates a violation of area-law. The black hole entanglement entropy saturates to a finite value (black line). The linear size in (b), (c) and (d) is $L=101$.}
\label{fig:EE}
\end{figure} 

\section{AdS/CFT correspondence}

We will first summarise how the AdS$_3$/CFT$_2$ correspondence can theoretically determine the entanglement entropy of a 3D BTZ black hole from the corresponding 2D boundary CFT. Then we will employ our black hole simulator to calculate the entanglement entropy across the horizon and identify the parametric regime where it agrees with the CFT prediction~\cite{ryu_aspects_2006,ryu_holographic_2006}.

In the holographic context, the Ryu--Takayanagi formula suggests that the entanglement entropy of a region $A$ with a length $\xi$ on the boundary $\cft_2$ is given by the area of the minimal surface $\gamma$ in the $\ads_3$ spacetime that is attached to the two endpoints of region $A$~\cite{ryu_aspects_2006,ryu_holographic_2006}, as shown in Fig. \ref{fig:EE}(a). In the presence of a black hole in $\ads_3$ spacetime, this holographic duality yields the entanglement entropy
\be
S^{\cft}(\beta,\xi)=\frac{c}{3}\ln\left[\frac{\beta}{\pi\epsilon} \sinh\frac{\pi \xi}{\beta}\right],
\label{eq:allS}
\ee
where $\epsilon$ denotes the UV cutoff, $c$ is the central charge and $\beta$ is inverse temperature \cite{ryu_aspects_2006,ryu_holographic_2006,Rangamani_2017}. Note that \eqref{eq:allS} has the same expression as the entanglement entropy of a thermal 2D CFT \cite{calabrese_entanglement_2004}. We now specialize in the case of a BTZ black hole with a large mass, $M$. In this semiclassical limit, the minimal path, $\gamma$, is shown in Fig. \ref{fig:EE}(a). This path gives the bipartition of both the CFT in $A$ that wraps around the whole space and its trivial complement $B$ as well as of the black hole where $\cal{A}$ is the outside of the black hole and $\cal{B}$ is inside. As the CFT bipartition is trivial, including the whole boundary, the entropy, $S^{\cft}$, is purely thermal. On the other hand, the black hole entropy, $S^{\rm BTZ}$, probes the quantum correlations of its pure ground state across the horizon. We now fix the boundary temperature to the BTZ Hawking temperature, $\beta=T_H^{-1}$, and use the Brown-Henneaux holographic formula, $c=3l/(2G)$, \cite{Brown1986} that relates the bulk properties of the black hole with the central charge of the boundary. Taking the length scale of the boundary to be $\xi\sim 2\pi l$, we find that the thermal 2D CFT entanglement is given by $S^{\cft}(T_H^{-1},2\pi l)$ \cite{Susskind1998,Azeyanagi2008,Cadoni2010}. 

We will now numerically determine the entanglement entropy of the BTZ black hole, $\sbh$, from the black hole simulator. To that end we construct the correlation matrix $\mathcal{C}$ with elements the two-point correlation functions $\mathcal{C}_{ij}=\langle \Phi | c_i^\dagger c_j | \Phi \rangle$, where $|\Phi\rangle$ is a many-body ground state of the Hamiltonian~\eqref{eq:Ham} and $i,j$ run through subsystem $\mathcal{B}$. Then the entanglement entropy between $\mathcal{B}$ and $\mathcal{A}$ is given by
\be
\sbh =-\sum_k \zeta_k \log(\zeta_k) + (1-\zeta_k) \log(1-\zeta_k),
\label{eq:entropy}
\ee
where the $\zeta_k$ are the positive eigenvalues of $\mathcal{C}$ \cite{latorre_short_2009,Guo2021}. The leading term in the resulting entanglement entropy of the black hole is expected to satisfy area law behavior. For $D=3$ the ``area" law takes the form
\begin{equation}
S(r_h)= k 2\pi r_h,
\label{eq:bh}
\end{equation}
where $2\pi r_h$ is the perimeter of the horizon. The constant $k=1/(4\Geff)$ can be expressed in terms of effective Newton constant $\Geff$ when the $S(r_h)$ is interpreted as the Bekenstein-Hawking entropy \cite{Srednicki1993,Jacobson1994,Susskind1994,Cadoni2007b,Satz2013,Zhou_2020}. We navigate around debates concerning both the species problem and the regularization problem of entanglement entropy by subsuming both issues within the definition of $\Geff$ \cite{Emparan:2006aa,solodukhin:2011aa}. Notably, we can manipulate $\Geff$ by modifying the regularization, or more precisely, the lattice spacing in the model. As demonstrated in Appendix \ref{app:entanglement}, $\Geff$ is directly proportional to the lattice spacing. In Fig.~\ref{fig:EE}(b), we show that entanglement entropy obtained from our simulator is given by an area law as in Eq.~\eqref{eq:bh}, with an effective Newton constant $\Geff\sim 0.7$ when $L=101$. In Fig.~\ref{fig:EE}(c), we see that the entanglement entropy of the 3D BTZ black hole simulator determined numerically from \eqref{eq:entropy} and the entanglement entropy of the corresponding boundary $\cft_2$, dictated by \eqref{eq:allS}, align remarkably well in the semiclassical limit $r_h\gg l$, where \eqref{eq:Ham} is valid. This agreement is shown quantitatively in Fig.~\ref{fig:EE}(c) either by increasing the radius $r_h$ for fixed curvature $l$ or by decreasing the curvature $l$ for fixed $r_h$. 

Note that Hamiltonian \eqref{eq:Ham} describes massless free fermions, and thus it is critical. Indeed, for a fixed value of the radius, we find that the entanglement entropy of flat spacetime, encoded in the simulator by uniform couplings, $F_i=F$, scales logarithmically with system size, as shown in Fig.~\ref{fig:EE}(d). On the other hand, the entanglement entropy across the horizon of a black hole stabilizes with system size to a finite-non-zero value, as shown in Fig.~\ref{fig:EE}(d). This ensures that the black hole entropy retains its "area" law behavior, unlike the flat case that depends on the system size.

\begin{figure}
\includegraphics[width=\linewidth]{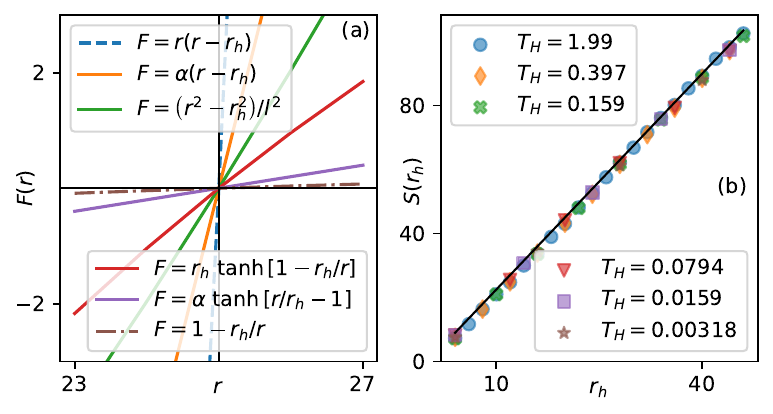}
\caption{(a) Different lapse functions with horizon at $r_h=20$ indicated by vertical black line. (b) All entanglement entropies are perfectly aligned regardless of their Hawking temperature ($\alpha=l=5$, $r_h=25$). The average slope of different lapse functions gives $\Geff\sim 0.7$ with a standard deviation given by $0.008$ for $L=101$. Black solid line shows the average slope.}
\label{fig:universality}
\end{figure}

\section{Entanglement entropy of various black holes}

We now consider several lapse functions each corresponding to different black hole profiles, see Fig.~\ref{fig:universality}(a). While they all have different Hawking temperatures they produce the same area law behavior in their entanglement entropy as the BTZ black hole, see Fig.~\ref{fig:universality}(b). Obtaining the same entropy for different black holes is known as the \emph{problem of universality} \cite{Carlip2002,Carlip:2007aa,Strominger:2009aa}. Such behavior is expected for BTZ black holes with different temperatures. Indeed, in the semiclassical limit, the entanglement entropy \eqref{eq:allS} is given by $S^{\cft} \propto c l / \beta$ for $\beta\ll l$. The $\ads/\cft$ relates $r_h \propto l^2 / \beta$ and via the Brown--Henneaux formula $c$ is proportional to $l/G$. Thus the entropy is independent of any particular characteristics of the black hole, such as its Hawking temperature.

Such an argument cannot be directly generalized when non-BTZ black holes are considered. Nevertheless, our simulator \eqref{eq:Ham} can explain this universal behavior for all black hole profiles parameterised by overall constants, in the following way. Overall factors in the lapse function $F(r)$ of the black hole geometry become also an overall factor in the simulator Hamiltonian \eqref{eq:Ham}. Since the two-point correlation matrix is invariant to such overall factors, the entanglement entropy stays the same for different black hole profiles even if they have different Hawking temperatures. Hence, any black hole profile that has negligible nonlinear terms around the horizon compared to the lattice spacing e.g., the ones considered in Fig.~\ref{fig:universality}(a), can be described by the same thermal CFT as the BTZ black hole. 

\section{Conclusions}

We have shown that our simulator is able to probe the quantum correlation properties of black holes. We observed that a whole set of 3D black holes have the same entanglement entropy as the one predicted by the $\cft_2$ dual to the BTZ black hole. Our results are in line with the interpretation of the Bekenstein-Hawking entropy as topological entanglement entropy \cite{McGough:2013aa}. Indeed, \eqref{eq:allS} indicates that the universal term comes from additive part, $S^{\rm top}=c \ln[\sinh(\pi \xi/\beta)]/3$, without UV cutoff. This topological term describes the thermal entropy of black hole in the semiclassical limit. 

Our universal black hole simulator is given in terms of free fermions that is analytically tractable making it viable to theoretical investigations, while it can be readily realized in the laboratory \cite{Munoz-de-Nova:2019aa,Yanay:2020aa}. Moreover, it can be directly applied also to higher dimensions, thus offering a simple and versatile medium to probe more complex questions, such as investigating the effect of black hole geometry on interacting fermions.

{\bf \em Acknowledgments:--}
We are grateful to Giandomenico Palumbo and Patricio Salgado-Rebolledo for helpful discussions. This work was in part supported by EPSRC Grant No. EP/R020612/1.

\appendix

\section{Dirac equation to lattice representation \label{app:dirac}}

The 3D Dirac equation in a rotationally symmetric geometry and away from the origin, $r\gg 1$ reduces to a 2D equation involving time and the radial coordinate $\slashed{\nabla}_\text{2D} \psi_\text{2D} \approx 0$. For 2D massless Dirac fermions the most general spinor takes the form $\psi_\text{2D} =[\phi(t,r),-\phi(t,r)]^T/\sqrt{2}$. Hence, the Dirac equation becomes
\begin{equation}
\partial_t \phi(t,r) = -\big[\partial_r (\tilde{F}(r) \phi(t,r)) + \tilde{F}(r) \partial_r \phi(t,r)\big]/4.
\label{seq:2Dirac1}
\end{equation}
The dependence on the angular coordinate $\theta$ has been suppressed as we are interested in the $r\gg 1$ limit, i.e. away from the origin (see the derivation of (3) in main text). 

The representation of \eqref{seq:2Dirac1} on a fermionic lattice is obtained by discretizing the spatial position with a lattice constant $a$ (we fix $a=1$). For simplicity, we consider a square lattice with fermionic tunneling couplings that depend on the radial distance $r$ in order to reproduce \eqref{seq:2Dirac1}. Note that as the contributions from the angular part are suppressed away from the origin, there is freedom in choosing its ``angular" couplings in the direction perpendicular to the ``radial" one.
To be more concrete we introduce lattice indices $(j,k)$. If $r$ is parallel to the $x$-axis then $\partial \phi_{j,k} \approx [\phi_{j+1,k}-\phi_{j-1,k}]/2$ that creates a contribution of tunneling in the $x$-direction with is the radial direction. If we add an angular contribution ($y$-direction) of tunneling then the derivative becomes $\partial \phi_{j,k} \approx [\phi_{j+1,k}-\phi_{j-1,k}+A(\phi_{j,k+1}-\phi_{j,k-1})]/2$ and the kinetic term will have an angular dependence that will be negligible in the $r\gg1$ limit. To make the couplings compatible with the rotational symmetry of the black hole geometry, we choose $A=1$ throughout the lattice, which makes the kinetic term contributions locally symmetric along the $x$ and $y$ directions. 
Moreover, note that as the propagation is along a square lattice we adopt the {\em Manhattan} distance rather than the Euclidian one. This change in distance measure deforms the geometry of the black hole away from the $x$ or $y$ axis without changing its thermalization properties nor its correlations across the horizon as we numerically verified. 

Using the product rule for finite difference formula
\be
\begin{split}
\partial (\tilde{F}_{j,k} \phi_{j,k}) \approx [\tilde{F}_{j+1,k}~\phi_{j+1,k}-\tilde{F}_{j-1,k}~\phi_{j-1,k}\\+\tilde{F}_{j,k+1}~\phi_{j,k+1}-\tilde{F}_{j,k-1}~\phi_{j,k-1}]/2,
\end{split}
\ee
Eq.~\eqref{seq:2Dirac1} becomes
\be
\dot{\phi}_{j,k} = -F_{j,k} \left[\phi_{j+1,k}-\phi_{j-1,k}+\phi_{j,k+1}-\phi_{j,k-1}\right]/4,
\ee
where for slowly varying $\tilde{F}_{j,k}$ we use $F_{j,k} \approx [\tilde{F}_{j,k}+\tilde{F}_{n}]/2$ where $\tilde{F}_{n}$ are nearest neighbours on lattice. Next, substituting $\phi_{j,k}=(-i)^j (-i)^k \hat{f}_{j,k}$, where $\hat{f}$ obeys Fermionic canonical commutation relation $\{\hat{f}_j,\hat{f}_k^\dagger\}=\delta_{jk}$, Dirac equation becomes
\be
i \dot{\hat{f}}_{j,k} = -F_{j,k} \left[\hat{f}_{j+1,k}+\hat{f}_{j-1,k}+\hat{f}_{j,k+1}+\hat{f}_{j,k-1}\right]/4.
\label{sep:dirac2}
\ee
Making use of Heisenberg equation of motion $i \dot{\hat{f}}_{j,k} =[\hat{f}_{j,k},\mathcal{H}]$, \eqref{sep:dirac2}, the resulting lattice system is described by free fermions on a two-dimensional square lattice with nearest neighboring hopping
\be
\begin{split}
\mathcal{H}=-\frac{1}{4}\sum_{j,k}F_{j,k} \Big[\hat{f}_{j,k}^\dagger \hat{f}_{j,k+1} + \hat{f}_{j,k}^\dagger \hat{f}_{j+1,k} \\+ \hat{f}_{j,k}^\dagger \hat{f}_{j,k-1} + \hat{f}_{j,k}^\dagger \hat{f}_{j-1,k} + {\rm h.c}\Big].
\end{split}
\ee
This can be written in compact form as
\be
\mathcal{H}=-\frac{1}{4}\sum_{\langle i,j \rangle}F_i \left(\hat{f}_{i}^\dagger \hat{f}_{j} + {\rm h.c}\right),
\label{eq:Ham}
\ee
thus reaching the Hamiltonian shown in the main text.

\begin{figure}
\centering
\includegraphics[width=\linewidth]{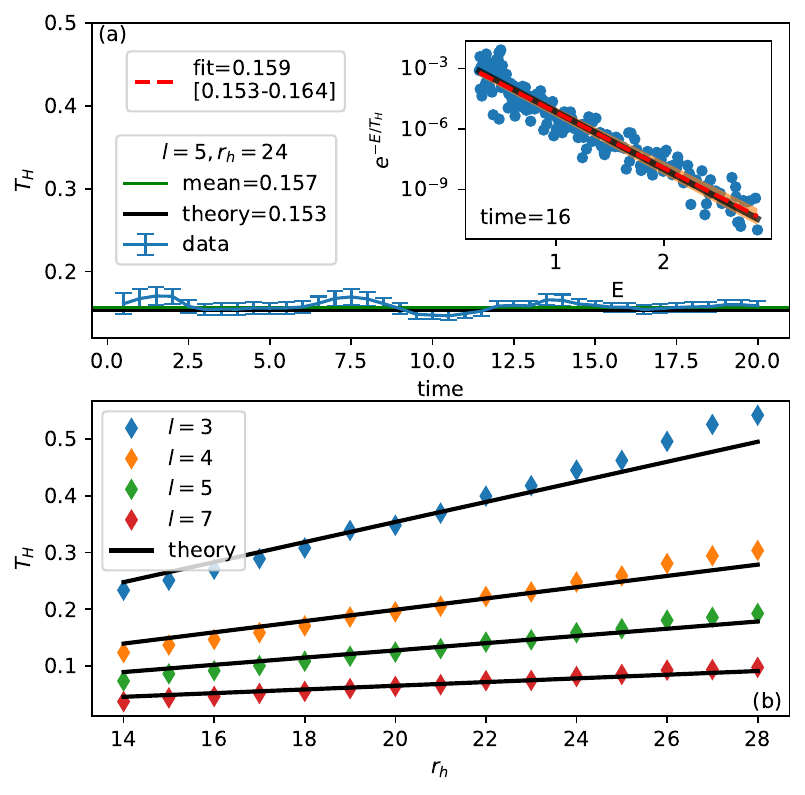}
\caption{Hawking temperature of 3D BTZ black hole for $L=61$ and error analysis. (a) Hawking temperature as a function of time. The inset figure shows a semi-log plot of energy modes located outside of the horizon. The slope gives the Hawking temperature. The margin of error in the inset is given in the bracket. (b) Hawking temperature is calculated for different horizon radiuses $r_h$ and cosmological constants $l^{-2}$. Only the weighted mean over t=0.2 to t=20 is shown.
A single Dirac particle is initialized in a superposition of four points at $t=0$ (blue crosses) just behind the black hole horizon with radius $r_h=10$ on a lattice with linear size $L=61$, $\eta=10$ and $\alpha=10^{-3}$. The dispersion of particle density is depicted at $t=1$. (b) The particle population that escaped the black hole appears as Hawking radiation at $t=8$. (c) Hawking radiation has thermal distribution. The slope on semi-log plot yields a Hawking temperature $T_H$ which is in good agreement with the theoretically predicted value, for $r_h=15$, $\eta=10$ and $\alpha=10^{-3}$. (d) Time-averaged Hawking temperatures over $t=\{0.2,8.0\}$ are depicted for a range of parameters $\eta$. The error bars indicate the standard deviation around the mean. Good agreement with the expected Hawking temperature $T_H=\eta/(4\pi)$ is obtained apart from large values of $\eta$, due to the finite lattice spacing, and for small $\eta$, due to finite size effects.
}
\label{fig:TH_BTZ}
\end{figure}
\section{Hawking temperature of 3D BTZ black hole \label{app:hawking}}

 The most celebrated quantum property of black holes is that quantum fluctuations escape their gravitational attraction. These fluctuations are witnessed outside the black hole as thermal radiation with temperature $T_H$ that depends on the geometrical characteristics of the black hole, as Hawking famously predicted in 1974 \cite{hawking_black_1974}. We now demonstrate that the fermionic lattice \eqref{eq:Ham} accurately describes 3D Dirac fermions in black hole geometry by determining the temperature of the escaped radiation. 

Consider for concreteness the BTZ black hole with profile $F=(r^2-r_h^2)/l^2$ where $l$ is related to the cosmological constant. Hawking temperature is given by
\be
T_H=\frac{1}{4\pi}\partial_r F (r_h)=\sqrt{2 G M}/(l\pi)
\ee
where $G$ is Newton constant and $M$ is mass of the black hole related to event horizon with $MG=r_h^2/(8l^2)$. Thus for given $r_h$ and $l$ we can obtain the mass of the black hole. To investigate the Hawking radiation with our lattice model we initially prepare a wave packet $|\psi(0) \rangle$ inside the black hole and monitor its quenched evolution as it escapes through the horizon. In particular, we initialize a single-particle state $|\psi(0)\rangle=\sum_{\{n\}}\lambda_{\{n\}} c_{\{n\}}^{\dagger}|0\rangle$ in an equal superposition on the $\{n\}$ sites on the inner region of the black hole horizon. Subsequently, we let the system evolve in time and we measure the probability density of the particle that is \emph{emitted} outside the black hole across the horizon at a given time $t$. Most of the population remains trapped inside the black hole \cite{Flouris2018} until eventually some \emph{escapes}, via quantum tunneling \cite{parikh_hawking_2000} through the horizon and moves to infinity.

The component of the wave packet outside the black hole corresponds to Hawking radiation if the population $P(E)=|\langle E|\psi(t)\rangle|^2$ of modes $|E\rangle$ with energy E that are the eigenstates of the Hamiltonian in the outer region. It is then expected that the Hawking radiation takes the thermal form $P(E)\propto e^{-E/T_H}$, where $T_H$ denotes the Hawking temperature. We numerically evolve the wave packet $|\psi(t)\rangle$ and calculate the corresponding Hawking temperature from a slope in a semi-log plot, as shown in Fig.~\ref{fig:TH_BTZ}(a). 

We find that the numerical Hawking temperature averaged over early times has an error of $3\%$. At last, in Fig.~\ref{fig:TH_BTZ}(b) we consider different $l$ values over a range of horizons and find good agreement between the numerical and theoretical values of the Hawking temperature.

\section{Unruh temperature of 3D BTZ black hole \label{app:unruh}}

In $(2+1)$D, the Schwarzschild metric is given by
\begin{equation}
ds^2 = f(r)dt^2 - \frac{1}{f(r)}dr^2 - r^2 d\theta^2,
\end{equation}
where $f(r)$ is some function such that $f(r_h) = 0$ and changes sign as we move across $r_h$, where $r_h$ is the location of the event horizon. Close to the horizon to the first order we have
\begin{equation}
f(r) \approx f(r_h) + (r-r_h)f'(r_h) \equiv  k (r-r_h).
\end{equation}
Therefore, the metric close to the horizon is given by
\begin{equation}
ds^2 = k(r-r_h) dt^2 - \frac{dr^2}{k(r-r_h)}  - r^2 d\theta^2.
\end{equation}
Let us now define the new coordinate $R^2 = k(r-r_h)$. Using this coordinate, the metric is 
\begin{equation}
ds^2  = R^2 dt^2 - \left(\frac{2}{k}\right)^2 dR^2 - \left( \frac{R^2}{k} + r_h \right)^2 d\theta^2.
\end{equation}
This looks very close to the Rindler metric describing a uniformly accelerating observer moving through a Minkowski spacetime. Let us make the final coordinate transformation $ \rho = 2R/k$, which gives us
\begin{equation}
ds^2 = \left( \frac{k \rho}{2} \right)^2 dt^2 - d \rho^2 - \left( \frac{k \rho^2 }{4}  + r_h \right)^2 d \theta^2.
\end{equation}
If we simplify this metric further by assuming that $\rho^2 \ll r_h$, then we arrive at the metric 
\begin{equation}
ds^2 = \left( \alpha \rho  \right)^2 dt^2 - d \rho^2 - r_h^2 d \theta^2, \label{eq:rindler_metric}
\end{equation}
where we have defined $\alpha = k/2 = f'(r_h)/2$.

We are interested in the Dirac field on this background and the Hawking radiation generated by it. In order to derive this, we note that this metric looks like the metric for the space-time $M = \mathbb{R}^2 \times S^1$, where $\mathbb{R}^2$ is a flat $(1+1)$D space-time endowed with the Rindler metric with acceleration $\alpha $, and $S^1$ is a circle of radius $r_h$. Therefore, we expect the Dirac field to exhibit the Unruh effect here and the angular portion to play no role for large $r_h$. In this coordinate system, the massless Dirac equation reads
\begin{equation}
\slashed{\nabla}_\mathrm{2D} \psi + \frac{1}{r_h} \partial_\theta \psi = 0,
\end{equation}
where $\slashed{\nabla} = e^\mu_a \gamma^a \partial_\mu + {1 \over 2\sqrt{|g|}} \gamma^a \partial_\mu(\sqrt{|g|} e^\mu_a) $ and $\slashed{\nabla}_\mathrm{2D}$ is simply the case with the $(1+1)$D Rindler metric substituted in. As the system is rotationally symmetric, we take the ansatz solution $\psi(t,\rho,\theta) = e^{-im \theta} \phi(t,\rho)$, where $\phi$ is a two-component spinor field. This yields the equation of motion 
\begin{equation}
\slashed{\nabla}_\mathrm{2D} \phi + \frac{i m }{r_h} \phi = 0
\end{equation}
For large $r_h$ and small angular momentum $m$, we arrive at
\begin{equation}
\slashed{\nabla}_\mathrm{2D} \phi  = 0
\end{equation}
so the non-trivial dynamics of the field is governed by the  Dirac equation on a Rindler metric. Using the chiral gamma matrix representation $\gamma^0 = i \sigma^x$ and $\gamma^1 = \sigma^y$, where $\sigma^i$ are the Pauli matrices, the (unnormalized) positive energy solutions are given by
\begin{equation}
\psi_{k,m} = u_k \frac{|\rho|^{ i k/\alpha}}{\sqrt{|\rho|}}e^{im \theta}, \ u_k = \begin{cases}
u_+ & \text{for $k \geq 0$} \\
u_- & \text{for $k < 0$}
\end{cases}, \label{eq:3d_wf}
\end{equation}
where $u_\pm$ are the two component eigenvectors of $\sigma^z$ where $\sigma^z u_\pm = \pm u_\pm$~\cite{Mertens2022}. The negative energy solutions are simply given by the complex conjugates. These solutions are only valid for $\rho > 0$ as they exist only in a single Rindler wedge.

As the Unruh effect requires us to measure the ground state of the Minkowski spacetime from the perspective of the Rindler observer, we must also have possession of the Minkowsi modes. The metric near the horizon can be written as
\begin{equation}
ds^2 = dt^2 - dX^2 - r_h^2 d\theta^2
\end{equation}
where the relationship between the coordinates is given by $T = \rho \sinh(\alpha t)$ and $X = \rho \cosh(\alpha t)$. The unnormalized positive energy solutions (using the same gamma matrix representation) on this metric are given by
\begin{equation}
\Psi_{k,m} =  u_k e^{i k X} e^{i m \theta} \label{eq:mink_wf}
\end{equation}
where $u_k$ is the same as defined in Eq.~(\ref{eq:3d_wf}) and $N$ is a normalisation constant. The negative energy solutions are obtained from the complex conjugate. Note that these solutions are valid for all $X$ so extend to the other side of the Rindler wedge. 

Let $a_{p,n}$ and $b_{p,n}$ be the particle and anti-particle modes of the Rindler observer associated with the solutions Eq.~(\ref{eq:3d_wf}), and let $A_{p,n}$ and $B_{p,n}$ be analogous for the Minkowski observer. The Minkowski observer defines their vacuum state (or ground state) as the state $|0_\mathrm{M}\rangle$ such that $A_{p,n}|0_\mathrm{M}\rangle = B_{p,n}|0_\mathrm{M}\rangle = 0$ for all $p$ and $n$. On the other hand, this state will not be the vacuum for the Rindler modes which is the source of the Unruh effect. Noting that our quantum field can be expressed with respect to either the Rindler modes of Eq.~(\ref{eq:3d_wf}) or the Minkowski modes of Eq.~(\ref{eq:mink_wf}), then this induces a Bogololiubov transformation of their corresponding mode operators allowing us to relate the Rindler and Minkowski mode operators linearly as
\begin{equation}
a_{k,m}  = \sum_n \int dq \left[ A_{q,n} (\psi_{k,m},\Psi_{q,m}) + B^\dagger_{q,n} (\psi_{k,m},\Psi^*_{q,n}) \right]    
\end{equation}
where 
\begin{equation}
( \psi, \phi ) = \int_0^\infty d\rho \int_0^{2\pi} r_h d\theta   \psi^\dagger \phi
\end{equation}
is the standard inner product for spinors on the spatial hypersurface induced by the metric of Eq.~(\ref{eq:rindler_metric}). Note that in order to perform this inner product between Minkowski and Rindler modes one must express both in the same coordinate system. Using the calculations of Ref.~\cite{Mertens2022}, the mode occupation of the Rindler modes in the Minkowski vacuum is given by
\begin{equation}
\langle 0_\mathrm{M}| a^\dagger_{p,m} a_{q,n} |0_\mathrm{M}\rangle = \frac{1}{e^{E_p/T} + 1} \delta_{mn} \delta(p-q) 
\end{equation}
where $T = \alpha/2\pi = f'(r_h)/4\pi$.

The previous calculation is exact in the Rindler frame, however, note that the Rindler frame exists only close to the horizon. The Dirac modes of the black hole frame will extend far from the horizon, however, we note that these modes reproduce the Hawking/Unruh effect well. In order to simulate this numerically on the lattice, we require two ingredients: the Minkowski vacuum $|0_\mathrm{M}\rangle$ and the modes which diagonalize the Hamiltonian in the Schwarzschild frame $a_{p,m}$. The vacuum $|0_\mathrm{M}\rangle$ is obtained easily as the many-body ground state of a homogeneous 2D lattice Hamiltonian. We then generate the Hamiltonian of the Schwarzschild frame and diagonalize it numerically to find its modes $a_{p,n}$. Then, one can calculate $\langle 0_\mathrm{M}| a^\dagger_p a_p |0_\mathrm{M}\rangle$ with possession of the correlation matrix of the model. This effect will only work for low energies as we are approximating a continuum effect with the lattice.

Note that the Minkowski Hamiltonian exists throughout the lattice, whereas the Schwarzschild Hamiltonian only has support outside of the horizon. This fact is the source of the thermality: probing the Minkowski modes with modes that exist only outside the horizon effectively performs the trace $\mathrm{tr}(|0_\mathrm{M}\rangle \langle 0_\mathrm{M}|) = e^{-\beta \mathcal{H}_\mathrm{ent}}$, where $\mathcal{H}_\mathrm{ent}$ is the entanglement Hamiltonian. The fact that the modes in the Schwarzschild frame produce a thermal spectrum implies the interesting observation that the entanglement Hamiltonian must be approximately equal to the Schwarzschild frame Hamiltonian, which was discussed in Ref.~\cite{Mertens2022}

\begin{figure}
\centering
\includegraphics[width=\linewidth]{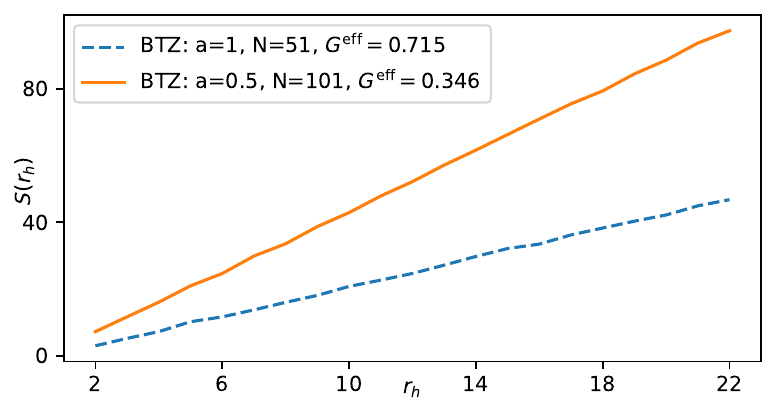}
\caption{Entanglement entropy of {\rm BTZ} ($l=5$) black hole for $L=51$. Different lattice spacings ($a$) shows that $\Geff$ decreases as $\Geff \propto a$.}
\label{fig:supp-lattice}
\end{figure}

\section{Entanglement entropy lattice regularisation \label{app:entanglement}}

As we are probing the quantum properties of the Dirac field the lattice regularisation influences the resulting entropy, $S(r_h)$. To that end, we consider the system to be of linear size $L$ and discretizing space with lattice spacing $a=L/N$ where $N$ is the number of lattice points within $L$. If we fix $r_h$ and $L$ and we increase $N$ then we obtain that $\Geff \propto a$, i.e. it goes to zero as $N$ increases. If we fix $N$ and $L$, i.e. fix the lattice spacing, then we obtain a fixed value for the gravitational constant $\Geff$. Subsequently, we change the radius $r_h$ to recover the area law dependence of the entanglement entropy $S_A \propto r_h/L\propto N$. In the main text, we choose $a=1$ and $N=101$ which results in $\Geff \approx 0.7$. In Fig.~\ref{fig:supp-lattice}, we consider other lattice spacing values and show that $S$ diverges and $\Geff$ decreases with decreasing lattice spacing.

\bibliography{ref}

\begin{thebibliography}{66}%
\makeatletter
\providecommand \@ifxundefined [1]{%
 \@ifx{#1\undefined}
}%
\providecommand \@ifnum [1]{%
 \ifnum #1\expandafter \@firstoftwo
 \else \expandafter \@secondoftwo
 \fi
}%
\providecommand \@ifx [1]{%
 \ifx #1\expandafter \@firstoftwo
 \else \expandafter \@secondoftwo
 \fi
}%
\providecommand \natexlab [1]{#1}%
\providecommand \enquote  [1]{``#1''}%
\providecommand \bibnamefont  [1]{#1}%
\providecommand \bibfnamefont [1]{#1}%
\providecommand \citenamefont [1]{#1}%
\providecommand \href@noop [0]{\@secondoftwo}%
\providecommand \href [0]{\begingroup \@sanitize@url \@href}%
\providecommand \@href[1]{\@@startlink{#1}\@@href}%
\providecommand \@@href[1]{\endgroup#1\@@endlink}%
\providecommand \@sanitize@url [0]{\catcode `\\12\catcode `\$12\catcode
  `\&12\catcode `\#12\catcode `\^12\catcode `\_12\catcode `\%12\relax}%
\providecommand \@@startlink[1]{}%
\providecommand \@@endlink[0]{}%
\providecommand \url  [0]{\begingroup\@sanitize@url \@url }%
\providecommand \@url [1]{\endgroup\@href {#1}{\urlprefix }}%
\providecommand \urlprefix  [0]{URL }%
\providecommand \Eprint [0]{\href }%
\providecommand \doibase [0]{https://doi.org/}%
\providecommand \selectlanguage [0]{\@gobble}%
\providecommand \bibinfo  [0]{\@secondoftwo}%
\providecommand \bibfield  [0]{\@secondoftwo}%
\providecommand \translation [1]{[#1]}%
\providecommand \BibitemOpen [0]{}%
\providecommand \bibitemStop [0]{}%
\providecommand \bibitemNoStop [0]{.\EOS\space}%
\providecommand \EOS [0]{\spacefactor3000\relax}%
\providecommand \BibitemShut  [1]{\csname bibitem#1\endcsname}%
\let\auto@bib@innerbib\@empty
\bibitem [{\citenamefont {Bekenstein}(1973)}]{Bekenstein1973}%
  \BibitemOpen
  \bibfield  {author} {\bibinfo {author} {\bibfnamefont {J.~D.}\ \bibnamefont
  {Bekenstein}},\ }\bibfield  {title} {\bibinfo {title} {Black holes and
  entropy},\ }\href {https://doi.org/10.1103/PhysRevD.7.2333} {\bibfield
  {journal} {\bibinfo  {journal} {Phys. Rev. D}\ }\textbf {\bibinfo {volume}
  {7}},\ \bibinfo {pages} {2333} (\bibinfo {year} {1973})}\BibitemShut
  {NoStop}%
\bibitem [{\citenamefont {Hawking}(1974)}]{hawking_black_1974}%
  \BibitemOpen
  \bibfield  {author} {\bibinfo {author} {\bibfnamefont {S.~W.}\ \bibnamefont
  {Hawking}},\ }\bibfield  {title} {\bibinfo {title} {Black hole explosions?},\
  }\href@noop {} {\bibfield  {journal} {\bibinfo  {journal} {Nature}\ }\textbf
  {\bibinfo {volume} {248}},\ \bibinfo {pages} {30} (\bibinfo {year}
  {1974})}\BibitemShut {NoStop}%
\bibitem [{\citenamefont {Hawking}(1975)}]{Hawking:1975aa}%
  \BibitemOpen
  \bibfield  {author} {\bibinfo {author} {\bibfnamefont {S.~W.}\ \bibnamefont
  {Hawking}},\ }\bibfield  {title} {\bibinfo {title} {Particle creation by
  black holes},\ }\href {https://doi.org/10.1007/BF02345020} {\bibfield
  {journal} {\bibinfo  {journal} {Communications in Mathematical Physics}\
  }\textbf {\bibinfo {volume} {43}},\ \bibinfo {pages} {199} (\bibinfo {year}
  {1975})}\BibitemShut {NoStop}%
\bibitem [{\citenamefont {Susskind}(2006)}]{Susskind:2006aa}%
  \BibitemOpen
  \bibfield  {author} {\bibinfo {author} {\bibfnamefont {L.}~\bibnamefont
  {Susskind}},\ }\bibfield  {title} {\bibinfo {title} {The paradox of quantum
  black holes},\ }\href {https://doi.org/10.1038/nphys429} {\bibfield
  {journal} {\bibinfo  {journal} {Nature Physics}\ }\textbf {\bibinfo {volume}
  {2}},\ \bibinfo {pages} {665} (\bibinfo {year} {2006})}\BibitemShut {NoStop}%
\bibitem [{\citenamefont {Garay}\ \emph {et~al.}(2000)\citenamefont {Garay},
  \citenamefont {Anglin}, \citenamefont {Cirac},\ and\ \citenamefont
  {Zoller}}]{Garay2000}%
  \BibitemOpen
  \bibfield  {author} {\bibinfo {author} {\bibfnamefont {L.~J.}\ \bibnamefont
  {Garay}}, \bibinfo {author} {\bibfnamefont {J.~R.}\ \bibnamefont {Anglin}},
  \bibinfo {author} {\bibfnamefont {J.~I.}\ \bibnamefont {Cirac}},\ and\
  \bibinfo {author} {\bibfnamefont {P.}~\bibnamefont {Zoller}},\ }\bibfield
  {title} {\bibinfo {title} {{Sonic Analog of Gravitational Black Holes in
  Bose-Einstein Condensates}},\ }\href
  {https://doi.org/10.1103/PhysRevLett.85.4643} {\bibfield  {journal} {\bibinfo
   {journal} {Phys. Rev. Lett.}\ }\textbf {\bibinfo {volume} {85}},\ \bibinfo
  {pages} {4643} (\bibinfo {year} {2000})}\BibitemShut {NoStop}%
\bibitem [{\citenamefont {Stone}(2013)}]{stone_analogue_2013}%
  \BibitemOpen
  \bibfield  {author} {\bibinfo {author} {\bibfnamefont {M.}~\bibnamefont
  {Stone}},\ }\bibfield  {title} {\bibinfo {title} {An analogue of {Hawking}
  radiation in the quantum {Hall} effect},\ }\href
  {https://doi.org/10.1088/0264-9381/30/8/085003} {\bibfield  {journal}
  {\bibinfo  {journal} {Classical and Quantum Gravity}\ }\textbf {\bibinfo
  {volume} {30}},\ \bibinfo {pages} {085003} (\bibinfo {year}
  {2013})}\BibitemShut {NoStop}%
\bibitem [{\citenamefont {Volovik}(2016)}]{volovik_black_2016}%
  \BibitemOpen
  \bibfield  {author} {\bibinfo {author} {\bibfnamefont {G.~E.}\ \bibnamefont
  {Volovik}},\ }\bibfield  {title} {\bibinfo {title} {Black hole and {Hawking}
  radiation by type-{{II Weyl}} fermions},\ }\href@noop {} {\bibfield
  {journal} {\bibinfo  {journal} {Jetp Lett.}\ }\textbf {\bibinfo {volume}
  {104}},\ \bibinfo {pages} {645} (\bibinfo {year} {2016})}\BibitemShut
  {NoStop}%
\bibitem [{\citenamefont {Lapierre}\ \emph {et~al.}(2020)\citenamefont
  {Lapierre}, \citenamefont {Choo}, \citenamefont {Tauber}, \citenamefont
  {Tiwari}, \citenamefont {Neupert},\ and\ \citenamefont
  {Chitra}}]{lapierre_emergent_2020}%
  \BibitemOpen
  \bibfield  {author} {\bibinfo {author} {\bibfnamefont {B.}~\bibnamefont
  {Lapierre}}, \bibinfo {author} {\bibfnamefont {K.}~\bibnamefont {Choo}},
  \bibinfo {author} {\bibfnamefont {C.}~\bibnamefont {Tauber}}, \bibinfo
  {author} {\bibfnamefont {A.}~\bibnamefont {Tiwari}}, \bibinfo {author}
  {\bibfnamefont {T.}~\bibnamefont {Neupert}},\ and\ \bibinfo {author}
  {\bibfnamefont {R.}~\bibnamefont {Chitra}},\ }\bibfield  {title} {\bibinfo
  {title} {Emergent black hole dynamics in critical floquet systems},\ }\href
  {https://doi.org/10.1103/PhysRevResearch.2.023085} {\bibfield  {journal}
  {\bibinfo  {journal} {Phys. Rev. Research}\ }\textbf {\bibinfo {volume}
  {2}},\ \bibinfo {pages} {023085} (\bibinfo {year} {2020})}\BibitemShut
  {NoStop}%
\bibitem [{\citenamefont {Rold\'an-Molina}\ \emph {et~al.}(2017)\citenamefont
  {Rold\'an-Molina}, \citenamefont {Nunez},\ and\ \citenamefont
  {Duine}}]{Magnonic2017}%
  \BibitemOpen
  \bibfield  {author} {\bibinfo {author} {\bibfnamefont {A.}~\bibnamefont
  {Rold\'an-Molina}}, \bibinfo {author} {\bibfnamefont {A.~S.}\ \bibnamefont
  {Nunez}},\ and\ \bibinfo {author} {\bibfnamefont {R.~A.}\ \bibnamefont
  {Duine}},\ }\bibfield  {title} {\bibinfo {title} {{Magnonic Black Holes}},\
  }\href {https://doi.org/10.1103/PhysRevLett.118.061301} {\bibfield  {journal}
  {\bibinfo  {journal} {Phys. Rev. Lett.}\ }\textbf {\bibinfo {volume} {118}},\
  \bibinfo {pages} {061301} (\bibinfo {year} {2017})}\BibitemShut {NoStop}%
\bibitem [{\citenamefont {Horner}\ \emph {et~al.}(2023)\citenamefont {Horner},
  \citenamefont {Hallam},\ and\ \citenamefont {Pachos}}]{Horner2023}%
  \BibitemOpen
  \bibfield  {author} {\bibinfo {author} {\bibfnamefont {M.~D.}\ \bibnamefont
  {Horner}}, \bibinfo {author} {\bibfnamefont {A.}~\bibnamefont {Hallam}},\
  and\ \bibinfo {author} {\bibfnamefont {J.~K.}\ \bibnamefont {Pachos}},\
  }\bibfield  {title} {\bibinfo {title} {Chiral spin-chain interfaces
  exhibiting event-horizon physics},\ }\href
  {https://doi.org/10.1103/PhysRevLett.130.016701} {\bibfield  {journal}
  {\bibinfo  {journal} {Phys. Rev. Lett.}\ }\textbf {\bibinfo {volume} {130}},\
  \bibinfo {pages} {016701} (\bibinfo {year} {2023})}\BibitemShut {NoStop}%
\bibitem [{\citenamefont {Bombelli}\ \emph {et~al.}(1986)\citenamefont
  {Bombelli}, \citenamefont {Koul}, \citenamefont {Lee},\ and\ \citenamefont
  {Sorkin}}]{Bombelli1986}%
  \BibitemOpen
  \bibfield  {author} {\bibinfo {author} {\bibfnamefont {L.}~\bibnamefont
  {Bombelli}}, \bibinfo {author} {\bibfnamefont {R.~K.}\ \bibnamefont {Koul}},
  \bibinfo {author} {\bibfnamefont {J.}~\bibnamefont {Lee}},\ and\ \bibinfo
  {author} {\bibfnamefont {R.~D.}\ \bibnamefont {Sorkin}},\ }\bibfield  {title}
  {\bibinfo {title} {Quantum source of entropy for black holes},\ }\href
  {https://doi.org/10.1103/PhysRevD.34.373} {\bibfield  {journal} {\bibinfo
  {journal} {Phys. Rev. D}\ }\textbf {\bibinfo {volume} {34}},\ \bibinfo
  {pages} {373} (\bibinfo {year} {1986})}\BibitemShut {NoStop}%
\bibitem [{\citenamefont {Srednicki}(1993)}]{Srednicki1993}%
  \BibitemOpen
  \bibfield  {author} {\bibinfo {author} {\bibfnamefont {M.}~\bibnamefont
  {Srednicki}},\ }\bibfield  {title} {\bibinfo {title} {{Entropy and area}},\
  }\href {https://doi.org/10.1103/PhysRevLett.71.666} {\bibfield  {journal}
  {\bibinfo  {journal} {Phys. Rev. Lett.}\ }\textbf {\bibinfo {volume} {71}},\
  \bibinfo {pages} {666} (\bibinfo {year} {1993})}\BibitemShut {NoStop}%
\bibitem [{\citenamefont {Cardy}(1986)}]{Cardy1986}%
  \BibitemOpen
  \bibfield  {author} {\bibinfo {author} {\bibfnamefont {J.~L.}\ \bibnamefont
  {Cardy}},\ }\bibfield  {title} {\bibinfo {title} {Operator content of
  two-dimensional conformally invariant theories},\ }\href
  {https://doi.org/https://doi.org/10.1016/0550-3213(86)90552-3} {\bibfield
  {journal} {\bibinfo  {journal} {Nuclear Physics B}\ }\textbf {\bibinfo
  {volume} {270}},\ \bibinfo {pages} {186} (\bibinfo {year}
  {1986})}\BibitemShut {NoStop}%
\bibitem [{\citenamefont {Kabat}\ and\ \citenamefont
  {Strassler}(1994)}]{kabat:1994aa}%
  \BibitemOpen
  \bibfield  {author} {\bibinfo {author} {\bibfnamefont {D.}~\bibnamefont
  {Kabat}}\ and\ \bibinfo {author} {\bibfnamefont {M.~J.}\ \bibnamefont
  {Strassler}},\ }\bibfield  {title} {\bibinfo {title} {A comment on entropy
  and area},\ }\href@noop {} {\bibfield  {journal} {\bibinfo  {journal}
  {Physics Letters B}\ }\textbf {\bibinfo {volume} {329}},\ \bibinfo {pages}
  {46} (\bibinfo {year} {1994})}\BibitemShut {NoStop}%
\bibitem [{\citenamefont {Strominger}\ and\ \citenamefont
  {Vafa}(1996)}]{Strominger1996}%
  \BibitemOpen
  \bibfield  {author} {\bibinfo {author} {\bibfnamefont {A.}~\bibnamefont
  {Strominger}}\ and\ \bibinfo {author} {\bibfnamefont {C.}~\bibnamefont
  {Vafa}},\ }\bibfield  {title} {\bibinfo {title} {Microscopic origin of the
  {Bekenstein-Hawking} entropy},\ }\href
  {https://doi.org/https://doi.org/10.1016/0370-2693(96)00345-0} {\bibfield
  {journal} {\bibinfo  {journal} {Physics Letters B}\ }\textbf {\bibinfo
  {volume} {379}},\ \bibinfo {pages} {99} (\bibinfo {year} {1996})}\BibitemShut
  {NoStop}%
\bibitem [{\citenamefont {Strominger}(1998)}]{Strominger1998}%
  \BibitemOpen
  \bibfield  {author} {\bibinfo {author} {\bibfnamefont {A.}~\bibnamefont
  {Strominger}},\ }\bibfield  {title} {\bibinfo {title} {Black hole entropy
  from near-horizon microstates},\ }\href
  {https://doi.org/10.1088/1126-6708/1998/02/009} {\bibfield  {journal}
  {\bibinfo  {journal} {Journal of High Energy Physics}\ }\textbf {\bibinfo
  {volume} {1998}},\ \bibinfo {pages} {009} (\bibinfo {year}
  {1998})}\BibitemShut {NoStop}%
\bibitem [{\citenamefont {Aharony}\ \emph {et~al.}(2000)\citenamefont
  {Aharony}, \citenamefont {Gubser}, \citenamefont {Maldacena}, \citenamefont
  {Ooguri},\ and\ \citenamefont {Oz}}]{Aharony:2000aa}%
  \BibitemOpen
  \bibfield  {author} {\bibinfo {author} {\bibfnamefont {O.}~\bibnamefont
  {Aharony}}, \bibinfo {author} {\bibfnamefont {S.~S.}\ \bibnamefont {Gubser}},
  \bibinfo {author} {\bibfnamefont {J.}~\bibnamefont {Maldacena}}, \bibinfo
  {author} {\bibfnamefont {H.}~\bibnamefont {Ooguri}},\ and\ \bibinfo {author}
  {\bibfnamefont {Y.}~\bibnamefont {Oz}},\ }\bibfield  {title} {\bibinfo
  {title} {Large {N} field theories, string theory and gravity},\ }\href
  {https://doi.org/https://doi.org/10.1016/S0370-1573(99)00083-6} {\bibfield
  {journal} {\bibinfo  {journal} {Physics Reports}\ }\textbf {\bibinfo {volume}
  {323}},\ \bibinfo {pages} {183} (\bibinfo {year} {2000})}\BibitemShut
  {NoStop}%
\bibitem [{\citenamefont {Emparan}(2006)}]{Emparan:2006aa}%
  \BibitemOpen
  \bibfield  {author} {\bibinfo {author} {\bibfnamefont {R.}~\bibnamefont
  {Emparan}},\ }\bibfield  {title} {\bibinfo {title} {Black hole entropy as
  entanglement entropy: a holographic derivation},\ }\href
  {https://doi.org/10.1088/1126-6708/2006/06/012} {\bibfield  {journal}
  {\bibinfo  {journal} {Journal of High Energy Physics}\ }\textbf {\bibinfo
  {volume} {2006}},\ \bibinfo {pages} {012} (\bibinfo {year}
  {2006})}\BibitemShut {NoStop}%
\bibitem [{\citenamefont {Maiella}\ and\ \citenamefont
  {Stornaiolo}(2007)}]{MAIELLA:2007aa}%
  \BibitemOpen
  \bibfield  {author} {\bibinfo {author} {\bibfnamefont {G.}~\bibnamefont
  {Maiella}}\ and\ \bibinfo {author} {\bibfnamefont {C.}~\bibnamefont
  {Stornaiolo}},\ }\bibfield  {title} {\bibinfo {title} {A {CFT} description of
  the {BTZ} black hole: Topology versus geometry (or thermodynamics versus
  statistical mechanics)},\ }\href {https://doi.org/10.1142/S0217751X07037111}
  {\bibfield  {journal} {\bibinfo  {journal} {International Journal of Modern
  Physics A}\ }\textbf {\bibinfo {volume} {22}},\ \bibinfo {pages} {3429}
  (\bibinfo {year} {2007})}\BibitemShut {NoStop}%
\bibitem [{\citenamefont {Solodukhin}(2011)}]{solodukhin:2011aa}%
  \BibitemOpen
  \bibfield  {author} {\bibinfo {author} {\bibfnamefont {S.~N.}\ \bibnamefont
  {Solodukhin}},\ }\bibfield  {title} {\bibinfo {title} {Entanglement entropy
  of black holes},\ }\href@noop {} {\bibfield  {journal} {\bibinfo  {journal}
  {Living Reviews in Relativity}\ }\textbf {\bibinfo {volume} {14}},\ \bibinfo
  {pages} {8} (\bibinfo {year} {2011})}\BibitemShut {NoStop}%
\bibitem [{\citenamefont {Ryu}\ and\ \citenamefont
  {Takayanagi}(2006{\natexlab{a}})}]{ryu_holographic_2006}%
  \BibitemOpen
  \bibfield  {author} {\bibinfo {author} {\bibfnamefont {S.}~\bibnamefont
  {Ryu}}\ and\ \bibinfo {author} {\bibfnamefont {T.}~\bibnamefont
  {Takayanagi}},\ }\bibfield  {title} {\bibinfo {title} {{Holographic
  Derivation of Entanglement Entropy from the anti--de Sitter Space/Conformal
  Field Theory Correspondence}},\ }\href
  {https://doi.org/10.1103/PhysRevLett.96.181602} {\bibfield  {journal}
  {\bibinfo  {journal} {Phys. Rev. Lett.}\ }\textbf {\bibinfo {volume} {96}},\
  \bibinfo {pages} {181602} (\bibinfo {year} {2006}{\natexlab{a}})}\BibitemShut
  {NoStop}%
\bibitem [{\citenamefont {Ryu}\ and\ \citenamefont
  {Takayanagi}(2006{\natexlab{b}})}]{ryu_aspects_2006}%
  \BibitemOpen
  \bibfield  {author} {\bibinfo {author} {\bibfnamefont {S.}~\bibnamefont
  {Ryu}}\ and\ \bibinfo {author} {\bibfnamefont {T.}~\bibnamefont
  {Takayanagi}},\ }\bibfield  {title} {\bibinfo {title} {Aspects of holographic
  entanglement entropy},\ }\href
  {https://doi.org/10.1088/1126-6708/2006/08/045} {\bibfield  {journal}
  {\bibinfo  {journal} {Journal of High Energy Physics}\ }\textbf {\bibinfo
  {volume} {2006}},\ \bibinfo {pages} {045} (\bibinfo {year}
  {2006}{\natexlab{b}})}\BibitemShut {NoStop}%
\bibitem [{\citenamefont {Rangamani}\ and\ \citenamefont
  {Takayanagi}(2017)}]{Rangamani_2017}%
  \BibitemOpen
  \bibfield  {author} {\bibinfo {author} {\bibfnamefont {M.}~\bibnamefont
  {Rangamani}}\ and\ \bibinfo {author} {\bibfnamefont {T.}~\bibnamefont
  {Takayanagi}},\ }\href {https://doi.org/10.1007/978-3-319-52573-0} {\emph
  {\bibinfo {title} {Holographic Entanglement Entropy}}}\ (\bibinfo
  {publisher} {Springer International Publishing},\ \bibinfo {year}
  {2017})\BibitemShut {NoStop}%
\bibitem [{\citenamefont {Almheiri}\ \emph {et~al.}(2021)\citenamefont
  {Almheiri}, \citenamefont {Hartman}, \citenamefont {Maldacena}, \citenamefont
  {Shaghoulian},\ and\ \citenamefont {Tajdini}}]{Almheiri2021}%
  \BibitemOpen
  \bibfield  {author} {\bibinfo {author} {\bibfnamefont {A.}~\bibnamefont
  {Almheiri}}, \bibinfo {author} {\bibfnamefont {T.}~\bibnamefont {Hartman}},
  \bibinfo {author} {\bibfnamefont {J.}~\bibnamefont {Maldacena}}, \bibinfo
  {author} {\bibfnamefont {E.}~\bibnamefont {Shaghoulian}},\ and\ \bibinfo
  {author} {\bibfnamefont {A.}~\bibnamefont {Tajdini}},\ }\bibfield  {title}
  {\bibinfo {title} {The entropy of {Hawking} radiation},\ }\href
  {https://doi.org/10.1103/RevModPhys.93.035002} {\bibfield  {journal}
  {\bibinfo  {journal} {Rev. Mod. Phys.}\ }\textbf {\bibinfo {volume} {93}},\
  \bibinfo {pages} {035002} (\bibinfo {year} {2021})}\BibitemShut {NoStop}%
\bibitem [{\citenamefont {Maldacena}(1999)}]{Maldacena1999}%
  \BibitemOpen
  \bibfield  {author} {\bibinfo {author} {\bibfnamefont {J.}~\bibnamefont
  {Maldacena}},\ }\bibfield  {title} {\bibinfo {title} {{The Large-N Limit of
  Superconformal Field Theories and Supergravity}},\ }\href
  {https://doi.org/10.1023/A:1026654312961} {\bibfield  {journal} {\bibinfo
  {journal} {International Journal of Theoretical Physics}\ }\textbf {\bibinfo
  {volume} {38}},\ \bibinfo {pages} {1113} (\bibinfo {year}
  {1999})}\BibitemShut {NoStop}%
\bibitem [{\citenamefont {'t~Hooft}(2009)}]{hooft:2009aa}%
  \BibitemOpen
  \bibfield  {author} {\bibinfo {author} {\bibfnamefont {G.}~\bibnamefont
  {'t~Hooft}},\ }\href@noop {} {\bibinfo {title} {Dimensional reduction in
  quantum gravity}} (\bibinfo {year} {2009}),\ \Eprint
  {https://arxiv.org/abs/gr-qc/9310026} {arXiv:gr-qc/9310026 [gr-qc]}
  \BibitemShut {NoStop}%
\bibitem [{\citenamefont {{\v C}ubrovi{\'c}}\ \emph {et~al.}(2009)\citenamefont
  {{\v C}ubrovi{\'c}}, \citenamefont {Zaanen},\ and\ \citenamefont
  {Schalm}}]{cubrovic:2009aa}%
  \BibitemOpen
  \bibfield  {author} {\bibinfo {author} {\bibfnamefont {M.}~\bibnamefont {{\v
  C}ubrovi{\'c}}}, \bibinfo {author} {\bibfnamefont {J.}~\bibnamefont
  {Zaanen}},\ and\ \bibinfo {author} {\bibfnamefont {K.}~\bibnamefont
  {Schalm}},\ }\bibfield  {title} {\bibinfo {title} {String theory, quantum
  phase transitions, and the emergent fermi liquid},\ }\href@noop {} {\bibfield
   {journal} {\bibinfo  {journal} {Science}\ }\textbf {\bibinfo {volume}
  {325}},\ \bibinfo {pages} {439} (\bibinfo {year} {2009})}\BibitemShut
  {NoStop}%
\bibitem [{\citenamefont {Liu}\ \emph {et~al.}(2011)\citenamefont {Liu},
  \citenamefont {McGreevy},\ and\ \citenamefont {Vegh}}]{liu:2011aa}%
  \BibitemOpen
  \bibfield  {author} {\bibinfo {author} {\bibfnamefont {H.}~\bibnamefont
  {Liu}}, \bibinfo {author} {\bibfnamefont {J.}~\bibnamefont {McGreevy}},\ and\
  \bibinfo {author} {\bibfnamefont {D.}~\bibnamefont {Vegh}},\ }\bibfield
  {title} {\bibinfo {title} {Non-fermi liquids from holography},\ }\href@noop
  {} {\bibfield  {journal} {\bibinfo  {journal} {Phys. Rev. D}\ }\textbf
  {\bibinfo {volume} {83}},\ \bibinfo {pages} {065029} (\bibinfo {year}
  {2011})}\BibitemShut {NoStop}%
\bibitem [{\citenamefont {Calabrese}\ and\ \citenamefont
  {Cardy}(2004)}]{calabrese_entanglement_2004}%
  \BibitemOpen
  \bibfield  {author} {\bibinfo {author} {\bibfnamefont {P.}~\bibnamefont
  {Calabrese}}\ and\ \bibinfo {author} {\bibfnamefont {J.}~\bibnamefont
  {Cardy}},\ }\bibfield  {title} {\bibinfo {title} {Entanglement entropy and
  quantum field theory},\ }\href
  {https://doi.org/10.1088/1742-5468/2004/06/p06002} {\bibfield  {journal}
  {\bibinfo  {journal} {J. Stat. Mech. Theory Exp.}\ }\textbf {\bibinfo
  {volume} {2004}},\ \bibinfo {pages} {P06002} (\bibinfo {year} {2004})},\
  \bibinfo {note} {publisher: IOP Publishing}\BibitemShut {NoStop}%
\bibitem [{\citenamefont {Eisert}\ \emph {et~al.}(2010)\citenamefont {Eisert},
  \citenamefont {Cramer},\ and\ \citenamefont
  {Plenio}}]{eisert_colloquium_2010}%
  \BibitemOpen
  \bibfield  {author} {\bibinfo {author} {\bibfnamefont {J.}~\bibnamefont
  {Eisert}}, \bibinfo {author} {\bibfnamefont {M.}~\bibnamefont {Cramer}},\
  and\ \bibinfo {author} {\bibfnamefont {M.~B.}\ \bibnamefont {Plenio}},\
  }\bibfield  {title} {\bibinfo {title} {Colloquium: Area laws for the
  entanglement entropy},\ }\href {https://doi.org/10.1103/RevModPhys.82.277}
  {\bibfield  {journal} {\bibinfo  {journal} {Rev. Mod. Phys.}\ }\textbf
  {\bibinfo {volume} {82}},\ \bibinfo {pages} {277} (\bibinfo {year}
  {2010})}\BibitemShut {NoStop}%
\bibitem [{\citenamefont {Casini}\ and\ \citenamefont
  {Huerta}(2005)}]{Casini_2005}%
  \BibitemOpen
  \bibfield  {author} {\bibinfo {author} {\bibfnamefont {H.}~\bibnamefont
  {Casini}}\ and\ \bibinfo {author} {\bibfnamefont {M.}~\bibnamefont
  {Huerta}},\ }\bibfield  {title} {\bibinfo {title} {Entanglement and alpha
  entropies for a massive scalar field in two dimensions},\ }\href
  {https://doi.org/10.1088/1742-5468/2005/12/P12012} {\bibfield  {journal}
  {\bibinfo  {journal} {Journal of Statistical Mechanics: Theory and
  Experiment}\ }\textbf {\bibinfo {volume} {2005}},\ \bibinfo {pages} {P12012}
  (\bibinfo {year} {2005})}\BibitemShut {NoStop}%
\bibitem [{\citenamefont {Nishioka}\ \emph {et~al.}(2009)\citenamefont
  {Nishioka}, \citenamefont {Ryu},\ and\ \citenamefont
  {Takayanagi}}]{nishioka_holographic_2009}%
  \BibitemOpen
  \bibfield  {author} {\bibinfo {author} {\bibfnamefont {T.}~\bibnamefont
  {Nishioka}}, \bibinfo {author} {\bibfnamefont {S.}~\bibnamefont {Ryu}},\ and\
  \bibinfo {author} {\bibfnamefont {T.}~\bibnamefont {Takayanagi}},\ }\bibfield
   {title} {\bibinfo {title} {Holographic entanglement entropy: an overview},\
  }\href {https://doi.org/10.1088/1751-8113/42/50/504008} {\bibfield  {journal}
  {\bibinfo  {journal} {Journal of Physics A: Mathematical and Theoretical}\
  }\textbf {\bibinfo {volume} {42}},\ \bibinfo {pages} {504008} (\bibinfo
  {year} {2009})}\BibitemShut {NoStop}%
\bibitem [{\citenamefont {Casini}\ \emph {et~al.}(2011)\citenamefont {Casini},
  \citenamefont {Huerta},\ and\ \citenamefont {Myers}}]{casini_towards_2011}%
  \BibitemOpen
  \bibfield  {author} {\bibinfo {author} {\bibfnamefont {H.}~\bibnamefont
  {Casini}}, \bibinfo {author} {\bibfnamefont {M.}~\bibnamefont {Huerta}},\
  and\ \bibinfo {author} {\bibfnamefont {R.~C.}\ \bibnamefont {Myers}},\
  }\bibfield  {title} {\bibinfo {title} {Towards a derivation of holographic
  entanglement entropy},\ }\href {https://doi.org/10.1007/JHEP05(2011)036}
  {\bibfield  {journal} {\bibinfo  {journal} {Journal of High Energy Physics}\
  }\textbf {\bibinfo {volume} {2011}},\ \bibinfo {pages} {36} (\bibinfo {year}
  {2011})}\BibitemShut {NoStop}%
\bibitem [{\citenamefont {Zhang}\ \emph {et~al.}(2011)\citenamefont {Zhang},
  \citenamefont {Grover},\ and\ \citenamefont
  {Vishwanath}}]{zhang_entanglement_2011}%
  \BibitemOpen
  \bibfield  {author} {\bibinfo {author} {\bibfnamefont {Y.}~\bibnamefont
  {Zhang}}, \bibinfo {author} {\bibfnamefont {T.}~\bibnamefont {Grover}},\ and\
  \bibinfo {author} {\bibfnamefont {A.}~\bibnamefont {Vishwanath}},\ }\bibfield
   {title} {\bibinfo {title} {{Entanglement Entropy of Critical Spin
  Liquids}},\ }\href {https://doi.org/10.1103/PhysRevLett.107.067202}
  {\bibfield  {journal} {\bibinfo  {journal} {Phys. Rev. Lett.}\ }\textbf
  {\bibinfo {volume} {107}},\ \bibinfo {pages} {067202} (\bibinfo {year}
  {2011})}\BibitemShut {NoStop}%
\bibitem [{\citenamefont {Morsink}\ and\ \citenamefont
  {Mann}(1991)}]{morsink_black_1991}%
  \BibitemOpen
  \bibfield  {author} {\bibinfo {author} {\bibfnamefont {S.~M.}\ \bibnamefont
  {Morsink}}\ and\ \bibinfo {author} {\bibfnamefont {R.~B.}\ \bibnamefont
  {Mann}},\ }\bibfield  {title} {\bibinfo {title} {Black hole radiation of
  dirac particles in 1+1 dimensions},\ }\href
  {https://doi.org/10.1088/0264-9381/8/12/010} {\bibfield  {journal} {\bibinfo
  {journal} {Classical and Quantum Gravity}\ }\textbf {\bibinfo {volume} {8}},\
  \bibinfo {pages} {2257} (\bibinfo {year} {1991})}\BibitemShut {NoStop}%
\bibitem [{\citenamefont {Mann}\ \emph {et~al.}(1991)\citenamefont {Mann},
  \citenamefont {Morsink}, \citenamefont {Sikkema},\ and\ \citenamefont
  {Steele}}]{Mann1991}%
  \BibitemOpen
  \bibfield  {author} {\bibinfo {author} {\bibfnamefont {R.~B.}\ \bibnamefont
  {Mann}}, \bibinfo {author} {\bibfnamefont {S.~M.}\ \bibnamefont {Morsink}},
  \bibinfo {author} {\bibfnamefont {A.~E.}\ \bibnamefont {Sikkema}},\ and\
  \bibinfo {author} {\bibfnamefont {T.~G.}\ \bibnamefont {Steele}},\ }\bibfield
   {title} {\bibinfo {title} {Semiclassical gravity in 1+1 dimensions},\ }\href
  {https://doi.org/10.1103/PhysRevD.43.3948} {\bibfield  {journal} {\bibinfo
  {journal} {Phys. Rev. D}\ }\textbf {\bibinfo {volume} {43}},\ \bibinfo
  {pages} {3948} (\bibinfo {year} {1991})}\BibitemShut {NoStop}%
\bibitem [{\citenamefont {Shi}\ \emph {et~al.}(2021)\citenamefont {Shi},
  \citenamefont {Yang}, \citenamefont {Xiang}, \citenamefont {Ge},
  \citenamefont {Li}, \citenamefont {Wang}, \citenamefont {Huang},
  \citenamefont {Tian}, \citenamefont {Song}, \citenamefont {Zheng},
  \citenamefont {Xu}, \citenamefont {Cai},\ and\ \citenamefont
  {Fan}}]{shi_chip_2021}%
  \BibitemOpen
  \bibfield  {author} {\bibinfo {author} {\bibfnamefont {Y.-H.}\ \bibnamefont
  {Shi}}, \bibinfo {author} {\bibfnamefont {R.-Q.}\ \bibnamefont {Yang}},
  \bibinfo {author} {\bibfnamefont {Z.}~\bibnamefont {Xiang}}, \bibinfo
  {author} {\bibfnamefont {Z.-Y.}\ \bibnamefont {Ge}}, \bibinfo {author}
  {\bibfnamefont {H.}~\bibnamefont {Li}}, \bibinfo {author} {\bibfnamefont
  {Y.-Y.}\ \bibnamefont {Wang}}, \bibinfo {author} {\bibfnamefont
  {K.}~\bibnamefont {Huang}}, \bibinfo {author} {\bibfnamefont
  {Y.}~\bibnamefont {Tian}}, \bibinfo {author} {\bibfnamefont {X.}~\bibnamefont
  {Song}}, \bibinfo {author} {\bibfnamefont {D.}~\bibnamefont {Zheng}},
  \bibinfo {author} {\bibfnamefont {K.}~\bibnamefont {Xu}}, \bibinfo {author}
  {\bibfnamefont {R.-G.}\ \bibnamefont {Cai}},\ and\ \bibinfo {author}
  {\bibfnamefont {H.}~\bibnamefont {Fan}},\ }\href@noop {} {\bibinfo {title}
  {On-chip black hole: {{Hawking}} radiation and curved spacetime in a
  superconducting quantum circuit with tunable couplers}} (\bibinfo {year}
  {2021}),\ \Eprint {https://arxiv.org/abs/2111.11092} {arXiv:2111.11092}
  \BibitemShut {NoStop}%
\bibitem [{\citenamefont {Unruh}(1976)}]{unruh1976}%
  \BibitemOpen
  \bibfield  {author} {\bibinfo {author} {\bibfnamefont {W.~G.}\ \bibnamefont
  {Unruh}},\ }\bibfield  {title} {\bibinfo {title} {Notes on black-hole
  evaporation},\ }\href@noop {} {\bibfield  {journal} {\bibinfo  {journal}
  {Phys. Rev. D}\ }\textbf {\bibinfo {volume} {14}},\ \bibinfo {pages} {870}
  (\bibinfo {year} {1976})}\BibitemShut {NoStop}%
\bibitem [{\citenamefont {Brown}\ and\ \citenamefont
  {Henneaux}(1986)}]{Brown1986}%
  \BibitemOpen
  \bibfield  {author} {\bibinfo {author} {\bibfnamefont {J.~D.}\ \bibnamefont
  {Brown}}\ and\ \bibinfo {author} {\bibfnamefont {M.}~\bibnamefont
  {Henneaux}},\ }\bibfield  {title} {\bibinfo {title} {Central charges in the
  canonical realization of asymptotic symmetries: An example from three
  dimensional gravity},\ }\href {https://doi.org/10.1007/BF01211590} {\bibfield
   {journal} {\bibinfo  {journal} {Communications in Mathematical Physics}\
  }\textbf {\bibinfo {volume} {104}},\ \bibinfo {pages} {207} (\bibinfo {year}
  {1986})}\BibitemShut {NoStop}%
\bibitem [{\citenamefont {Cadoni}\ and\ \citenamefont
  {Melis}(2010)}]{Cadoni2010}%
  \BibitemOpen
  \bibfield  {author} {\bibinfo {author} {\bibfnamefont {M.}~\bibnamefont
  {Cadoni}}\ and\ \bibinfo {author} {\bibfnamefont {M.}~\bibnamefont {Melis}},\
  }\bibfield  {title} {\bibinfo {title} {Holographic entanglement entropy of
  the {BTZ} black hole},\ }\href {https://doi.org/10.1007/s10701-010-9430-6}
  {\bibfield  {journal} {\bibinfo  {journal} {Foundations of Physics}\ }\textbf
  {\bibinfo {volume} {40}},\ \bibinfo {pages} {638} (\bibinfo {year}
  {2010})}\BibitemShut {NoStop}%
\bibitem [{\citenamefont {Khveshchenko}(2013)}]{Khveshchenko_2013}%
  \BibitemOpen
  \bibfield  {author} {\bibinfo {author} {\bibfnamefont {D.~V.}\ \bibnamefont
  {Khveshchenko}},\ }\bibfield  {title} {\bibinfo {title} {Simulating analogue
  holography in flexible dirac metals},\ }\href
  {https://doi.org/10.1209/0295-5075/104/47002} {\bibfield  {journal} {\bibinfo
   {journal} {Europhysics Letters}\ }\textbf {\bibinfo {volume} {104}},\
  \bibinfo {pages} {47002} (\bibinfo {year} {2013})}\BibitemShut {NoStop}%
\bibitem [{\citenamefont {Palumbo}\ and\ \citenamefont
  {Pachos}(2016)}]{Palumbo:2016aa}%
  \BibitemOpen
  \bibfield  {author} {\bibinfo {author} {\bibfnamefont {G.}~\bibnamefont
  {Palumbo}}\ and\ \bibinfo {author} {\bibfnamefont {J.~K.}\ \bibnamefont
  {Pachos}},\ }\bibfield  {title} {\bibinfo {title} {Holographic correspondence
  in topological superconductors},\ }\href
  {https://doi.org/https://doi.org/10.1016/j.aop.2016.05.005} {\bibfield
  {journal} {\bibinfo  {journal} {Annals of Physics}\ }\textbf {\bibinfo
  {volume} {372}},\ \bibinfo {pages} {175} (\bibinfo {year}
  {2016})}\BibitemShut {NoStop}%
\bibitem [{\citenamefont {Carlip}(2002)}]{Carlip2002}%
  \BibitemOpen
  \bibfield  {author} {\bibinfo {author} {\bibfnamefont {S.}~\bibnamefont
  {Carlip}},\ }\bibfield  {title} {\bibinfo {title} {{Near-Horizon Conformal
  Symmetry and Black Hole Entropy}},\ }\href
  {https://doi.org/10.1103/PhysRevLett.88.241301} {\bibfield  {journal}
  {\bibinfo  {journal} {Phys. Rev. Lett.}\ }\textbf {\bibinfo {volume} {88}},\
  \bibinfo {pages} {241301} (\bibinfo {year} {2002})}\BibitemShut {NoStop}%
\bibitem [{\citenamefont {Carlip}(2007)}]{Carlip:2007aa}%
  \BibitemOpen
  \bibfield  {author} {\bibinfo {author} {\bibfnamefont {S.}~\bibnamefont
  {Carlip}},\ }\bibfield  {title} {\bibinfo {title} {Black hole entropy and the
  problem of universality},\ }\href
  {https://doi.org/10.1088/1742-6596/67/1/012022} {\bibfield  {journal}
  {\bibinfo  {journal} {Journal of Physics: Conference Series}\ }\textbf
  {\bibinfo {volume} {67}},\ \bibinfo {pages} {012022} (\bibinfo {year}
  {2007})}\BibitemShut {NoStop}%
\bibitem [{\citenamefont {Strominger}(2009)}]{Strominger:2009aa}%
  \BibitemOpen
  \bibfield  {author} {\bibinfo {author} {\bibfnamefont {A.}~\bibnamefont
  {Strominger}},\ }\bibfield  {title} {\bibinfo {title} {Five problems in
  quantum gravity},\ }\href
  {https://doi.org/https://doi.org/10.1016/j.nuclphysbps.2009.07.049}
  {\bibfield  {journal} {\bibinfo  {journal} {Nuclear Physics B - Proceedings
  Supplements}\ }\textbf {\bibinfo {volume} {192-193}},\ \bibinfo {pages} {119}
  (\bibinfo {year} {2009})}\BibitemShut {NoStop}%
\bibitem [{\citenamefont {Mu{\~n}oz~de Nova}\ \emph {et~al.}(2019)\citenamefont
  {Mu{\~n}oz~de Nova}, \citenamefont {Golubkov}, \citenamefont {Kolobov},\ and\
  \citenamefont {Steinhauer}}]{Munoz-de-Nova:2019aa}%
  \BibitemOpen
  \bibfield  {author} {\bibinfo {author} {\bibfnamefont {J.~R.}\ \bibnamefont
  {Mu{\~n}oz~de Nova}}, \bibinfo {author} {\bibfnamefont {K.}~\bibnamefont
  {Golubkov}}, \bibinfo {author} {\bibfnamefont {V.~I.}\ \bibnamefont
  {Kolobov}},\ and\ \bibinfo {author} {\bibfnamefont {J.}~\bibnamefont
  {Steinhauer}},\ }\bibfield  {title} {\bibinfo {title} {Observation of thermal
  {Hawking} radiation and its temperature in an analogue black hole},\ }\href
  {https://doi.org/10.1038/s41586-019-1241-0} {\bibfield  {journal} {\bibinfo
  {journal} {Nature}\ }\textbf {\bibinfo {volume} {569}},\ \bibinfo {pages}
  {688} (\bibinfo {year} {2019})}\BibitemShut {NoStop}%
\bibitem [{\citenamefont {Yanay}\ \emph {et~al.}(2020)\citenamefont {Yanay},
  \citenamefont {Braum{\"u}ller}, \citenamefont {Gustavsson}, \citenamefont
  {Oliver},\ and\ \citenamefont {Tahan}}]{Yanay:2020aa}%
  \BibitemOpen
  \bibfield  {author} {\bibinfo {author} {\bibfnamefont {Y.}~\bibnamefont
  {Yanay}}, \bibinfo {author} {\bibfnamefont {J.}~\bibnamefont
  {Braum{\"u}ller}}, \bibinfo {author} {\bibfnamefont {S.}~\bibnamefont
  {Gustavsson}}, \bibinfo {author} {\bibfnamefont {W.~D.}\ \bibnamefont
  {Oliver}},\ and\ \bibinfo {author} {\bibfnamefont {C.}~\bibnamefont
  {Tahan}},\ }\bibfield  {title} {\bibinfo {title} {Two-dimensional hard-core
  {Bose--Hubbard} model with superconducting qubits},\ }\href
  {https://doi.org/10.1038/s41534-020-0269-1} {\bibfield  {journal} {\bibinfo
  {journal} {npj Quantum Information}\ }\textbf {\bibinfo {volume} {6}},\
  \bibinfo {pages} {58} (\bibinfo {year} {2020})}\BibitemShut {NoStop}%
\bibitem [{\citenamefont {{Lopez-Ortega}}(2009)}]{lopez-ortega_dirac_2009}%
  \BibitemOpen
  \bibfield  {author} {\bibinfo {author} {\bibfnamefont {A.}~\bibnamefont
  {{Lopez-Ortega}}},\ }\href@noop {} {\bibinfo {title} {The {{Dirac}} equation
  in {{D-dimensional}} spherically symmetric spacetimes}} (\bibinfo {year}
  {2009}),\ \Eprint {https://arxiv.org/abs/0906.2754} {arXiv:0906.2754 [gr-qc]}
  \BibitemShut {NoStop}%
\bibitem [{\citenamefont {Camporesi}\ and\ \citenamefont
  {Higuchi}(1996)}]{camporesi_eigenfunctions_1996}%
  \BibitemOpen
  \bibfield  {author} {\bibinfo {author} {\bibfnamefont {R.}~\bibnamefont
  {Camporesi}}\ and\ \bibinfo {author} {\bibfnamefont {A.}~\bibnamefont
  {Higuchi}},\ }\bibfield  {title} {\bibinfo {title} {On the eigenfunctions of
  the {{Dirac}} operator on spheres and real hyperbolic spaces},\ }\href@noop
  {} {\bibfield  {journal} {\bibinfo  {journal} {Journal of Geometry and
  Physics}\ }\textbf {\bibinfo {volume} {20}},\ \bibinfo {pages} {1} (\bibinfo
  {year} {1996})}\BibitemShut {NoStop}%
\bibitem [{\citenamefont {Yang}\ \emph {et~al.}(2020)\citenamefont {Yang},
  \citenamefont {Liu}, \citenamefont {Zhu}, \citenamefont {Luo},\ and\
  \citenamefont {Cai}}]{yang_simulating_2020}%
  \BibitemOpen
  \bibfield  {author} {\bibinfo {author} {\bibfnamefont {R.-Q.}\ \bibnamefont
  {Yang}}, \bibinfo {author} {\bibfnamefont {H.}~\bibnamefont {Liu}}, \bibinfo
  {author} {\bibfnamefont {S.}~\bibnamefont {Zhu}}, \bibinfo {author}
  {\bibfnamefont {L.}~\bibnamefont {Luo}},\ and\ \bibinfo {author}
  {\bibfnamefont {R.-G.}\ \bibnamefont {Cai}},\ }\bibfield  {title} {\bibinfo
  {title} {Simulating quantum field theory in curved spacetime with quantum
  many-body systems},\ }\href@noop {} {\bibfield  {journal} {\bibinfo
  {journal} {Phys. Rev. Research}\ }\textbf {\bibinfo {volume} {2}},\ \bibinfo
  {pages} {023107} (\bibinfo {year} {2020})}\BibitemShut {NoStop}%
\bibitem [{\citenamefont {Pedernales}\ \emph {et~al.}(2018)\citenamefont
  {Pedernales}, \citenamefont {Beau}, \citenamefont {Pittman}, \citenamefont
  {Egusquiza}, \citenamefont {Lamata}, \citenamefont {Solano},\ and\
  \citenamefont {{del Campo}}}]{pedernales_dirac_2018}%
  \BibitemOpen
  \bibfield  {author} {\bibinfo {author} {\bibfnamefont {J.~S.}\ \bibnamefont
  {Pedernales}}, \bibinfo {author} {\bibfnamefont {M.}~\bibnamefont {Beau}},
  \bibinfo {author} {\bibfnamefont {S.~M.}\ \bibnamefont {Pittman}}, \bibinfo
  {author} {\bibfnamefont {I.~L.}\ \bibnamefont {Egusquiza}}, \bibinfo {author}
  {\bibfnamefont {L.}~\bibnamefont {Lamata}}, \bibinfo {author} {\bibfnamefont
  {E.}~\bibnamefont {Solano}},\ and\ \bibinfo {author} {\bibfnamefont
  {A.}~\bibnamefont {{del Campo}}},\ }\bibfield  {title} {\bibinfo {title}
  {Dirac {{Equation}} in ( 1 + 1 )-{{Dimensional Curved Spacetime}} and the
  {{Multiphoton Quantum Rabi Model}}},\ }\href@noop {} {\bibfield  {journal}
  {\bibinfo  {journal} {Phys. Rev. Lett.}\ }\textbf {\bibinfo {volume} {120}},\
  \bibinfo {pages} {160403} (\bibinfo {year} {2018})}\BibitemShut {NoStop}%
\bibitem [{\citenamefont {Ba\~nados}\ \emph {et~al.}(1992)\citenamefont
  {Ba\~nados}, \citenamefont {Teitelboim},\ and\ \citenamefont
  {Zanelli}}]{BTZ1992}%
  \BibitemOpen
  \bibfield  {author} {\bibinfo {author} {\bibfnamefont {M.}~\bibnamefont
  {Ba\~nados}}, \bibinfo {author} {\bibfnamefont {C.}~\bibnamefont
  {Teitelboim}},\ and\ \bibinfo {author} {\bibfnamefont {J.}~\bibnamefont
  {Zanelli}},\ }\bibfield  {title} {\bibinfo {title} {{Black hole in
  three-dimensional spacetime}},\ }\href
  {https://doi.org/10.1103/PhysRevLett.69.1849} {\bibfield  {journal} {\bibinfo
   {journal} {Phys. Rev. Lett.}\ }\textbf {\bibinfo {volume} {69}},\ \bibinfo
  {pages} {1849} (\bibinfo {year} {1992})}\BibitemShut {NoStop}%
\bibitem [{\citenamefont {Mann}\ and\ \citenamefont
  {Solodukhin}(1997)}]{Mann1997}%
  \BibitemOpen
  \bibfield  {author} {\bibinfo {author} {\bibfnamefont {R.~B.}\ \bibnamefont
  {Mann}}\ and\ \bibinfo {author} {\bibfnamefont {S.~N.}\ \bibnamefont
  {Solodukhin}},\ }\bibfield  {title} {\bibinfo {title} {Quantum scalar field
  on a three-dimensional {(BTZ)} black hole instanton: Heat kernel, effective
  action, and thermodynamics},\ }\href
  {https://doi.org/10.1103/PhysRevD.55.3622} {\bibfield  {journal} {\bibinfo
  {journal} {Phys. Rev. D}\ }\textbf {\bibinfo {volume} {55}},\ \bibinfo
  {pages} {3622} (\bibinfo {year} {1997})}\BibitemShut {NoStop}%
\bibitem [{\citenamefont {Mertens}\ \emph {et~al.}(2022)\citenamefont
  {Mertens}, \citenamefont {Moghaddam}, \citenamefont {Chernyavsky},
  \citenamefont {Morice}, \citenamefont {van~den Brink},\ and\ \citenamefont
  {van Wezel}}]{Mertens2022}%
  \BibitemOpen
  \bibfield  {author} {\bibinfo {author} {\bibfnamefont {L.}~\bibnamefont
  {Mertens}}, \bibinfo {author} {\bibfnamefont {A.~G.}\ \bibnamefont
  {Moghaddam}}, \bibinfo {author} {\bibfnamefont {D.}~\bibnamefont
  {Chernyavsky}}, \bibinfo {author} {\bibfnamefont {C.}~\bibnamefont {Morice}},
  \bibinfo {author} {\bibfnamefont {J.}~\bibnamefont {van~den Brink}},\ and\
  \bibinfo {author} {\bibfnamefont {J.}~\bibnamefont {van Wezel}},\ }\bibfield
  {title} {\bibinfo {title} {Thermalization by a synthetic horizon},\
  }\href@noop {} {\bibfield  {journal} {\bibinfo  {journal} {Phys. Rev. Res.}\
  }\textbf {\bibinfo {volume} {4}},\ \bibinfo {pages} {043084} (\bibinfo {year}
  {2022})}\BibitemShut {NoStop}%
\bibitem [{\citenamefont {Susskind}\ and\ \citenamefont
  {Witten}(1998)}]{Susskind1998}%
  \BibitemOpen
  \bibfield  {author} {\bibinfo {author} {\bibfnamefont {L.}~\bibnamefont
  {Susskind}}\ and\ \bibinfo {author} {\bibfnamefont {E.}~\bibnamefont
  {Witten}},\ }\href {https://doi.org/10.48550/ARXIV.HEP-TH/9805114} {\bibinfo
  {title} {The holographic bound in {Anti-de Sitter} space}} (\bibinfo {year}
  {1998})\BibitemShut {NoStop}%
\bibitem [{\citenamefont {Azeyanagi}\ \emph {et~al.}(2008)\citenamefont
  {Azeyanagi}, \citenamefont {Nishioka},\ and\ \citenamefont
  {Takayanagi}}]{Azeyanagi2008}%
  \BibitemOpen
  \bibfield  {author} {\bibinfo {author} {\bibfnamefont {T.}~\bibnamefont
  {Azeyanagi}}, \bibinfo {author} {\bibfnamefont {T.}~\bibnamefont
  {Nishioka}},\ and\ \bibinfo {author} {\bibfnamefont {T.}~\bibnamefont
  {Takayanagi}},\ }\bibfield  {title} {\bibinfo {title} {Near extremal black
  hole entropy as entanglement entropy via
  {${\mathrm{AdS}}_{2}/{\mathrm{CFT}}_{1}$}},\ }\href
  {https://doi.org/10.1103/PhysRevD.77.064005} {\bibfield  {journal} {\bibinfo
  {journal} {Phys. Rev. D}\ }\textbf {\bibinfo {volume} {77}},\ \bibinfo
  {pages} {064005} (\bibinfo {year} {2008})}\BibitemShut {NoStop}%
\bibitem [{\citenamefont {Latorre}\ and\ \citenamefont
  {Riera}(2009)}]{latorre_short_2009}%
  \BibitemOpen
  \bibfield  {author} {\bibinfo {author} {\bibfnamefont {J.~I.}\ \bibnamefont
  {Latorre}}\ and\ \bibinfo {author} {\bibfnamefont {A.}~\bibnamefont
  {Riera}},\ }\bibfield  {title} {\bibinfo {title} {A short review on
  entanglement in quantum spin systems},\ }\href
  {https://doi.org/10.1088/1751-8113/42/50/504002} {\bibfield  {journal}
  {\bibinfo  {journal} {J. Phys. A}\ }\textbf {\bibinfo {volume} {42}},\
  \bibinfo {pages} {504002} (\bibinfo {year} {2009})},\ \bibinfo {note}
  {publisher: IOP Publishing}\BibitemShut {NoStop}%
\bibitem [{\citenamefont {Guo}\ \emph {et~al.}(2021)\citenamefont {Guo},
  \citenamefont {Yu}, \citenamefont {Huang}, \citenamefont {Yang},
  \citenamefont {Chi}, \citenamefont {Liao},\ and\ \citenamefont
  {Xiang}}]{Guo2021}%
  \BibitemOpen
  \bibfield  {author} {\bibinfo {author} {\bibfnamefont {Y.-B.}\ \bibnamefont
  {Guo}}, \bibinfo {author} {\bibfnamefont {Y.-C.}\ \bibnamefont {Yu}},
  \bibinfo {author} {\bibfnamefont {R.-Z.}\ \bibnamefont {Huang}}, \bibinfo
  {author} {\bibfnamefont {L.-P.}\ \bibnamefont {Yang}}, \bibinfo {author}
  {\bibfnamefont {R.-Z.}\ \bibnamefont {Chi}}, \bibinfo {author} {\bibfnamefont
  {H.-J.}\ \bibnamefont {Liao}},\ and\ \bibinfo {author} {\bibfnamefont
  {T.}~\bibnamefont {Xiang}},\ }\bibfield  {title} {\bibinfo {title}
  {Entanglement entropy of non-hermitian free fermions},\ }\href
  {https://doi.org/10.1088/1361-648X/ac216e} {\bibfield  {journal} {\bibinfo
  {journal} {Journal of Physics: Condensed Matter}\ }\textbf {\bibinfo {volume}
  {33}},\ \bibinfo {pages} {475502} (\bibinfo {year} {2021})}\BibitemShut
  {NoStop}%
\bibitem [{\citenamefont {Jacobson}(1994)}]{Jacobson1994}%
  \BibitemOpen
  \bibfield  {author} {\bibinfo {author} {\bibfnamefont {T.}~\bibnamefont
  {Jacobson}},\ }\href {https://doi.org/10.48550/ARXIV.GR-QC/9404039} {\bibinfo
  {title} {Black hole entropy and induced gravity}} (\bibinfo {year}
  {1994})\BibitemShut {NoStop}%
\bibitem [{\citenamefont {Susskind}\ and\ \citenamefont
  {Uglum}(1994)}]{Susskind1994}%
  \BibitemOpen
  \bibfield  {author} {\bibinfo {author} {\bibfnamefont {L.}~\bibnamefont
  {Susskind}}\ and\ \bibinfo {author} {\bibfnamefont {J.}~\bibnamefont
  {Uglum}},\ }\bibfield  {title} {\bibinfo {title} {Black hole entropy in
  canonical quantum gravity and superstring theory},\ }\href
  {https://doi.org/10.1103/PhysRevD.50.2700} {\bibfield  {journal} {\bibinfo
  {journal} {Phys. Rev. D}\ }\textbf {\bibinfo {volume} {50}},\ \bibinfo
  {pages} {2700} (\bibinfo {year} {1994})}\BibitemShut {NoStop}%
\bibitem [{\citenamefont {Cadoni}(2007)}]{Cadoni2007b}%
  \BibitemOpen
  \bibfield  {author} {\bibinfo {author} {\bibfnamefont {M.}~\bibnamefont
  {Cadoni}},\ }\bibfield  {title} {\bibinfo {title} {Induced gravity and
  entanglement entropy of {2D} black holes}\ }\href
  {https://doi.org/10.48550/arXiv.0709.0163} {10.48550/arXiv.0709.0163}
  (\bibinfo {year} {2007})\BibitemShut {NoStop}%
\bibitem [{\citenamefont {Satz}\ and\ \citenamefont
  {Jacobson}(2013)}]{Satz2013}%
  \BibitemOpen
  \bibfield  {author} {\bibinfo {author} {\bibfnamefont {A.}~\bibnamefont
  {Satz}}\ and\ \bibinfo {author} {\bibfnamefont {T.}~\bibnamefont
  {Jacobson}},\ }\href {https://doi.org/10.48550/ARXIV.1301.3171} {\bibinfo
  {title} {Black hole entropy and the renormalization group}} (\bibinfo {year}
  {2013})\BibitemShut {NoStop}%
\bibitem [{\citenamefont {Zhou}\ and\ \citenamefont {Kuang}(2020)}]{Zhou_2020}%
  \BibitemOpen
  \bibfield  {author} {\bibinfo {author} {\bibfnamefont {Y.-T.}\ \bibnamefont
  {Zhou}}\ and\ \bibinfo {author} {\bibfnamefont {X.-M.}\ \bibnamefont
  {Kuang}},\ }\bibfield  {title} {\bibinfo {title} {Scalar field in massive btz
  black hole and entanglement entropy},\ }\href@noop {} {\bibfield  {journal}
  {\bibinfo  {journal} {Chinese Physics C}\ }\textbf {\bibinfo {volume} {44}},\
  \bibinfo {pages} {015102} (\bibinfo {year} {2020})}\BibitemShut {NoStop}%
\bibitem [{\citenamefont {McGough}\ and\ \citenamefont
  {Verlinde}(2013)}]{McGough:2013aa}%
  \BibitemOpen
  \bibfield  {author} {\bibinfo {author} {\bibfnamefont {L.}~\bibnamefont
  {McGough}}\ and\ \bibinfo {author} {\bibfnamefont {H.}~\bibnamefont
  {Verlinde}},\ }\bibfield  {title} {\bibinfo {title} {{Bekenstein-Hawking}
  entropy as topological entanglement entropy},\ }\href
  {https://doi.org/10.1007/JHEP11(2013)208} {\bibfield  {journal} {\bibinfo
  {journal} {Journal of High Energy Physics}\ }\textbf {\bibinfo {volume}
  {2013}},\ \bibinfo {pages} {208} (\bibinfo {year} {2013})}\BibitemShut
  {NoStop}%
\bibitem [{\citenamefont {Flouris}\ \emph {et~al.}(2018)\citenamefont
  {Flouris}, \citenamefont {Mendoza~Jimenez}, \citenamefont {Debus},\ and\
  \citenamefont {Herrmann}}]{Flouris2018}%
  \BibitemOpen
  \bibfield  {author} {\bibinfo {author} {\bibfnamefont {K.}~\bibnamefont
  {Flouris}}, \bibinfo {author} {\bibfnamefont {M.}~\bibnamefont
  {Mendoza~Jimenez}}, \bibinfo {author} {\bibfnamefont {J.-D.}\ \bibnamefont
  {Debus}},\ and\ \bibinfo {author} {\bibfnamefont {H.~J.}\ \bibnamefont
  {Herrmann}},\ }\bibfield  {title} {\bibinfo {title} {Confining massless dirac
  particles in two-dimensional curved space},\ }\href
  {https://doi.org/10.1103/PhysRevB.98.155419} {\bibfield  {journal} {\bibinfo
  {journal} {Phys. Rev. B}\ }\textbf {\bibinfo {volume} {98}},\ \bibinfo
  {pages} {155419} (\bibinfo {year} {2018})}\BibitemShut {NoStop}%
\bibitem [{\citenamefont {Parikh}\ and\ \citenamefont
  {Wilczek}(2000)}]{parikh_hawking_2000}%
  \BibitemOpen
  \bibfield  {author} {\bibinfo {author} {\bibfnamefont {M.~K.}\ \bibnamefont
  {Parikh}}\ and\ \bibinfo {author} {\bibfnamefont {F.}~\bibnamefont
  {Wilczek}},\ }\bibfield  {title} {\bibinfo {title} {{Hawking Radiation as
  Tunneling}},\ }\href@noop {} {\bibfield  {journal} {\bibinfo  {journal}
  {Phys. Rev. Lett.}\ }\textbf {\bibinfo {volume} {85}},\ \bibinfo {pages}
  {5042} (\bibinfo {year} {2000})},\ \Eprint
  {https://arxiv.org/abs/hep-th/9907001} {arXiv:hep-th/9907001} \BibitemShut
  {NoStop}%
\end{thebibliography}%

\end{document}